\documentclass[12pt]{article}
\usepackage{amsmath}
\usepackage{graphicx}
\usepackage{enumerate}
\usepackage{natbib}
\usepackage{url} 
\usepackage{amsfonts}
\usepackage{xcolor}
\usepackage{float}

\newcommand{\blind}{1}

\newtheorem{theorem}{Theorem}

\begin{document}

\def\spacingset#1{\renewcommand{\baselinestretch}%
{#1}\small\normalsize} \spacingset{1}


\if1\blind
{
  \title{\bf Combining Covariate Adjustment with Group Sequential, Information Adaptive Designs to Improve Randomized Trial Efficiency}
  \author{
  Kelly Van Lancker, Joshua Betz and Michael Rosenblum\\
  Department of Biostatistics, Johns Hopkins Bloomberg School of Public Health, \\Baltimore, U.S.A.}
  \maketitle
} \fi

\if0\blind
{
  \bigskip
  \bigskip
  \bigskip
  \begin{center}
    {\LARGE\bf Combining Covariate Adjustment with Group Sequential, Information Adaptive Designs to Improve Randomized Trial Efficiency}
\end{center}
  \medskip
} \fi

\bigskip
\begin{abstract}
In clinical trials, there is potential to improve precision and reduce the required sample size by appropriately adjusting for baseline variables in the statistical analysis. This is called covariate adjustment. Despite recommendations by regulatory agencies in favor of covariate adjustment, it remains underutilized leading to inefficient trials. We address two obstacles that make it challenging to use covariate adjustment. A first obstacle is the incompatibility of many covariate adjusted estimators with commonly used boundaries in group sequential designs (GSDs). A second obstacle is the uncertainty at the design stage about how much precision gain will result from covariate adjustment. 
We propose a method that modifies the original estimator so that it becomes compatible with GSDs, while increasing or leaving unchanged the estimator's precision. Our approach allows the use of any asymptotically linear estimator, which covers many estimators used in randomized trials. Building on this, we propose using an information adaptive design, that is, continuing the trial until the required information level is achieved. Such a design adapts to the amount of precision gain and can lead to faster, more efficient trials, without sacrificing validity or power.
We evaluate estimator performance in simulations that mimic features of a completed stroke trial.
\end{abstract}

\noindent%
\vfill

\newpage

 \section{Introduction}\label{sec:intro}
In clinical trials, baseline data are collected on important participant characteristics  (e.g., age, baseline disease severity, and comorbidities).     
Covariate adjustment (i.e., adjusting for prespecified, prognostic baseline variables) is a statistical analysis method for estimating the average treatment effect that has high potential to improve precision for many trials \citep{tsiatis2008covariate, benkeser2020improving}. 

Despite the extensive literature on model-robust covariate adjustment methods \citep[e.g.,][]{koch1998issues, YangTsiatis2001, tsiatis2008covariate, Moore2009, Moore2009a, Zhang2015}
and recommendations by the U.S. Food and Drug Administration and the European Medicines Agency to use covariate adjustment when there are prognostic baseline variables \citep{FDA1998, FDA2020, FDA2021}, it remains highly underutilized. 
This is especially true for trials with binary, ordinal, and time-to-event outcomes, which are quite common in practice. This is problematic because the resulting analyses are inefficient by not fully exploiting the available information in the data, thereby forfeiting the opportunity to reduce the required sample size and/or trial duration. This can lead to unnecessary numbers of patients being exposed to an experimental treatment, which is unethical. We address two obstacles  that lead to this underutilization.

A first obstacle is the incompatibility of many covariate adjusted estimators with  commonly used stopping boundaries in group sequential designs (GSDs), when models used to construct the estimators are  misspecified. 
Specifically, to apply GSDs, the sequential test statistics need to have the independent increments covariance structure in order to control Type I error \citep{scharfstein1997semiparametric, jennison1997group, jennisonturnbull1999group}. Although these papers consider covariate adjusted estimators in GSDs, they restricted their theorems to estimators that are semiparametric efficient.
The general theory of \cite{scharfstein1997semiparametric} and \cite{jennison1997group} is not guaranteed to hold for covariate adjusted estimators under model misspecification, which is likely to be the case in practice. 
In particular, under model misspecification, covariate adjusted estimators can fail to have this independent increments property when using data of patients for whom the primary outcome is not measured yet. Since GSDs --especially with  \cite{o1979multiple} or  \cite{pocock1977group}  stopping boundaries-- are extensively used in practice for both efficiency and ethical reasons \citep{hatfield2016adaptive}, this incompatibility is an obstacle to realizing precision gains from covariate adjustment. We next describe our approach to tackling this obstacle.

We propose a general method that extends the highly useful theory of information-monitoring in GSDs  \citep{scharfstein1997semiparametric,jennison1997group}
 so that it can be used with  any regular,  asymptotically linear estimator. This covers many estimators in RCTs including the aforementioned covariate adjusted estimators, as well as other types of estimators 
\citep[e.g.,][]{jennison1991note, lee1992sequential, YangTsiatis2001,tsiatis2008covariate, zhang2008improving,qu2015estimation, diaz2019improved, benkeser2020improving}.
Also, our approach enables the use of different estimators at different stages of a trial.  This may be of interest, e.g.,  if one wants to use unadjusted estimators at earlier analysis times and  covariate adjusted estimators at later analysis times in a trial, as explained in Section~\ref{subsec:asympt}. 

Specifically, our method uses orthogonalization to produce modified estimators that (1) have the independent increments property needed to apply GSDs, and (2) simultaneously improve (or leave unchanged) the variance at each analysis. Such a method is needed in order to fully leverage prognostic baseline variables, leading to faster, more efficient trials for many disease areas, without sacrificing validity or power. 

A second obstacle to using covariate adjustment in practice is the uncertainty at the design stage about  the amount of precision gain and corresponding sample size reduction that should be expected from covariate adjustment. Proposals have been made  to use an external trial dataset to estimate the precision gain from using covariate adjusted estimators \citep[see e.g.,][]{li2021estimating}. Nevertheless, an incorrect projection of a covariate's prognostic value risks an over- or underpowered future trial.


To address this second obstacle, we  propose to use a trial design where  the analysis timing is based on accruing information
and is data adaptive \citep{scharfstein1997semiparametric,mehta2001flexible, tsiatis2006information, Zhang2009}. 
 In particular, we continuously monitor the accruing information (i.e., the reciprocal of the estimator's variance) during the trial and conduct interim analyses (and the final analysis) when  prespecified information thresholds are reached. We refer to  this type of design as ``information adaptive".
  Since adaptations to the analysis timing are made in a preplanned way based only on nuisance parameters, they are generally acceptable to regulators  \citep{FDA2019}.
 A special case of an information adaptive design is an event driven trial (commonly used in practice for time-to-event outcomes), where the trial continues  until  the required total number of events (which is approximately proportional to the information) has occurred \citep{freidlin2016information}; however, this typically uses an unadjusted estimator.
 
 To the best of our knowledge,  information adaptive designs have not been combined with covariate adjustment as we advocate here. This combination  leads to designs that  automatically adapt to the amount of precision gain due to covariate adjustment,  resulting in trials that are correctly powered.
 As covariate adjusted estimators typically have smaller variance (compared to corresponding unadjusted estimators), information will accrue faster and thus combining  covariate adjustment with 
 this information adaptive design will yield faster trials at no additional cost.


 In Section~\ref{sec:notation}, we introduce the data structure, estimands and estimators we will consider in the remainder of the article.
We describe in Section~\ref{sec:GSD} the proposed approach to combine covariate adjustment with GSDs. 
In Section~\ref{sec:inf}, we propose an information adaptive design. In Section~\ref{sec:dataAnal}, we demonstrate the performance of the proposed methods and design through simulation studies that mimic key features of the MISTIE III stroke trial \citep{hanley2019efficacy}. We end with a discussion in Section~\ref{sec:disc}.

\section{Data Structure, Estimands and Estimators}\label{sec:notation}
\subsection{Data Structure}
We consider the general data structure as described in \cite{scharfstein1997semiparametric} as it covers many common data structures used in randomized trials. The full data collected for each participant $i$ ($i=1, \dots n$) is described by the process  $X_i=\left\{E_i, (D_{u,i}, u\geq 0)\right\}$. Here, $E_i$ denotes the entry time into the study and $D_{u,i}$ all the additional data collected during the first $u$ time units on study. Throughout,  $D_{u,i}$ includes at least the study arm assignment $A_i$ (taking  value $0$ for the control arm and $1$ for the treatment arm) and a vector $W_i$ of baseline (i.e., pre-randomization) variables. Additionally, it includes a primary outcome and possibly other  post-baseline variables. 
This data structure can handle common outcomes such as continuous, binary, ordinal, and time-to-event. 
 We let $X=\left\{E, (D_{u}, u\geq 0)\right\}$ denote the full data for a generic participant.

As in \cite{scharfstein1997semiparametric}, we assume that data are collected in a fixed interval $[0,T]$ and that each participant's data $X_i$ are an independent and identically distributed (i.i.d.) draw from an unknown joint distribution $P_0$ on $X$.  The model on $P_0$ is nonparametric except  that (by design) we assume study arm $A$ is  assigned independent of 
 entry time and baseline variables $(E, W)$. Our asymptotics involve the sample size $n$ going to infinity while the time horizon $[0,T]$ is fixed.

The restriction of the full data for participant $i$ to the data collected up to any calendar time $t$ is represented by the data process $X_{t,i}=\left\{\Delta_i(t), \Delta_i(t)E_i, (D_{u,i}, 0\leq u\leq t-E_i)\right\}$, where $\Delta_i(t)$ is the indicator of already  having entered the study at time $t$ (i.e., $\Delta_i(t)=1$ if $E_i\leq t$ and $0$ otherwise). These data processes $X_{t,1}, \dots, X_{t,n}$ are also i.i.d. 

\subsection{Estimands and Estimators}
We consider arbitrary, real-valued  estimands, i.e.,  targets of inference. These are typically contrasts between summaries of the outcome distribution under assignment to treatment versus control, for the corresponding study population. We consider 
 estimands that represent marginal treatment effects, e.g.,  
the population risk difference, relative risk, or odds ratio for binary outcomes
, among others. The importance of clearly describing the estimand to be used in a clinical trial's primary analysis is stated in the ICH E9(R1) Addendum on estimands and sensitivity analyses \citep{ich2019}.
As in  \cite{scharfstein1997semiparametric}, we assume that the estimand (denoted $\theta$) is pathwise differentiable.

We next consider estimators (i.e., functions of the data) for each estimand. As in \cite{scharfstein1997semiparametric}, we assume that the estimators being considered  (denoted  $\widehat\theta$) are regular and  asymptotically linear (RAL) and consistent for $\theta$. A difference with their work is that we do not assume that our estimators are semiparametric efficient. In particular, all covariate adjusted estimators mentioned in Section~\ref{sec:intro} and also the covariate adjusted estimator for binary outcomes suggested in the recent FDA draft guidance on covariate adjustment \citep{FDA2021} may fail to be semiparametric efficient under model misspecification. Below we give a list of example {\it estimands} (in italics), each followed by a covariate adjusted estimator. The corresponding unadjusted estimators are the sample analog of the estimand. These can be obtained by replacing the population averages by sample means in the estimand definition. For example, for the difference in (population) mean outcomes estimand, the unadjusted estimator is the difference in sample means.
\begin{itemize}
  \item[] \textit{Difference in means of the primary outcome between study arms for continuous outcomes:} The ANCOVA (analysis of covariance) estimator can be used when the outcome is measured at a single time point \citep{YangTsiatis2001}. 
  This can be obtained by a two step procedure. First, one fits a linear regression model for the outcome given baseline variables, an indicator of study arm assignment, an intercept, and (optionally)  baseline variable by study arm assignment interactions. Second,  one uses standardization to compute the ANCOVA estimator; specifically, one first computes a predicted outcome under each possible arm assignment for each enrolled participant based on the fitted model, and then one takes the difference of the sample means over all participants (pooling across arms) of the predicted outcomes under $A=1$ and $A=0$. 
  For repeatedly measured continuous outcomes, the mixed-effects model for repeated measures (MMRM) estimator can be used to estimate the average treatment effect (i.e., difference in means) at the last visit \citep{wang2021model}.
  \item[] \textit{Risk difference for binary outcomes:} A covariate-adjusted estimator of this quantity can be obtained by following a similar standardization approach as for the continuous outcome but by replacing the linear regression model by a logistic regression model. A generalization of the covariate adjusted estimator in \cite{ge2011covariate}, is described in Appendix~\ref{sec:covAdj} of the Supplementary Materials. 
  Alternatively, we can focus on \textit{relative risk} or \textit{odds ratio} estimands by calculating respectively the ratio and the odds ratio based on the sample means over all participants (pooling across arms) of the predicted outcome under $A=1$ and $A=0$ \citep{Moore2009,benkeser2020improving}. 
    \item[] \textit{Log-odds ratio for ordinal outcomes:} 
    For an  outcome that is ordinal with levels $1, \dots, K$, 
    this estimand is the average of the cumulative log odds ratios over levels 1 to $K-1$  \citep{diaz2016enhanced, benkeser2020improving}. Model-robust, covariate adjusted estimators were proposed by \cite{diaz2016enhanced} and \cite{benkeser2020improving}. 
    Alternatively, these estimates of the arm-specific cumulative distribution functions can be used to estimate the \textit{Mann-Whitney estimand} \citep[see e.g.,][]{vermeulen2015increasing, benkeser2020improving}. This estimand reports the probability that a random patient assigned to the experimental treatment will have a better outcome than a random patient assigned to control, with ties broken at random. 
    \item[] \textit{Difference in restricted mean survival times for time-to-event outcomes:}  This estimand reports the expected value of a survival time that is truncated at a specified time $\tau$ \citep[see e.g.,][]{chen2001causal, royston2011use}. \cite{diaz2019improved} proposed a target minimum loss-based estimator, which is a model-robust, covariate adjusted estimator, for this estimand.
  Analogous to the other outcome cases, estimation involves first estimating the time-specific hazard conditional on baseline variables, and then marginalizing the corresponding survival probabilities (via transformation of the time-specific hazards using the product-limit formula) using the estimated covariate distribution pooled across arms \citep{diaz2019improved, benkeser2020improving}. \\
    A similar approach can be followed to estimate the \textit{survival probability difference} (i.e., difference between arm-specific probabilities of survival to a specified time point) or the \textit{relative risk} (i.e., ratio of the arm-specific probabilities of survival to a specified time point) \citep{benkeser2020improving}. 
\end{itemize}

\section{Orthogonalizing Estimators To Get  Independent \\Increments Property}\label{sec:GSD}
 Group sequential, information-based designs are described by \cite{scharfstein1997semiparametric, jennison1997group, jennisonturnbull1999group}. They entail analyzing the data at $K$ different analysis times $t_1, \dots, t_K$. We consider these as fixed times throughout this section, and handle the case of data dependent analysis times in Section~\ref{sec:inf}. Let $\theta$ denote the estimand.  
 At each analysis time $t_k$, $n$ independent draws (one for each  participant) of $X_{t_k}$ are available to test the null hypothesis $H_0: \theta= \theta_0$ against the sequence of local alternatives   $H_A:\theta_n = \theta_0 + \tau / \sqrt{n}$, for constant $\tau>0$.
 
 We next briefly discuss the motivation for using local alternatives, which were also used by \cite{scharfstein1997semiparametric}. According to  \cite{vaart_1998}, one uses a sequence of alternatives that converge to the null hypothesis at rate $O(1/\sqrt{n})$ because the corresponding testing problem is feasible (i.e., it's possible to achieve asymptotic power greater than the nominal significance level $\alpha$) but  non-trivial (i.e., it's not the case that all reasonable tests have power converging to 1).

For each analysis time  $t_k$, an  estimator $\widehat\theta_{t_k}$ and its corresponding  standardized (Wald) test statistic $Z_k=Z(t_k)=\left(\widehat\theta_{t_k}-\theta_0 \right)\left/ \widehat{se}(\widehat\theta_{t_k}) \right.$ are calculated, where 
$\widehat{se}(\widehat\theta_{t_k})$ denotes the estimated standard error of $\widehat\theta_{t_k}$.
The information accrued at the corresponding analysis time $t_k$ is defined as the reciprocal of the  estimator's variance, that is, $\widehat{\mathcal{I}}_k = (\widehat{se}(\widehat\theta_{t_k}))^{-2}$. 
Similar to \cite{scharfstein1997semiparametric} (Section 3, p. 1344), we assume that for each $k\leq K$,
\begin{equation}\lim_{n \rightarrow \infty}\widehat{\mathcal{I}}_{k}/n=\lim_{n \rightarrow \infty}\left\{n\widehat{Var}(\widehat\theta_{t_k})\right\}^{-1}=\lim_{n \rightarrow \infty}\left\{nVar(\widehat\theta_{t_k})\right\}^{-1}=\mathcal{I}_{k}^*>0,
\label{variance_convergence_assumption}
\end{equation}
where $\widehat{Var}(\widehat\theta_{t_k})$ denotes the estimated variance of $\widehat\theta_{t_k}$,  $Var(\widehat\theta_{t_k})$ the (true) variance of $\widehat\theta_{t_k}$,
and 
$\mathcal{I}_{k}^*$ is   called the inverse of the asymptotic variance of  $\widehat\theta_{t_k}$ (as $n \rightarrow \infty$).
In the above display, the first equality is by definition, the second represents convergence in probability (since the estimated variance is random), and the third represents convergence of a real-valued sequence to  the  finite limit $\mathcal{I}^*_k$.
 It follows from Section 3 of  \cite{scharfstein1997semiparametric} that $\mathcal{I}^*_k$ is less than or equal to the  semiparametric  information bound for estimating $\theta$ using the data up to time $t_k$ (with equality if $\widehat\theta_{t_k}$ is a semiparametric efficient estimator).

It follows from the above assumptions (including (\ref{variance_convergence_assumption}) and the estimators are RAL and consistent) that the vector of test statistics $(Z_1,\dots,Z_K)$ converges  in distribution to a multivariate normal with mean $\boldsymbol{\delta}$  and covariance matrix $\boldsymbol{\Sigma}$ under the null hypothesis (where $\boldsymbol{\delta}=\mathbf{0}$) and under the alternative hypothesis (where $\boldsymbol{\delta}=\tau \left(\sqrt{\mathcal{I}_{1}^*},\dots, \sqrt{\mathcal{I}_{K}^*}\right)$). We  assume that  $\boldsymbol{\Sigma}$  can consistently be  estimated by $\boldsymbol{\widehat\Sigma}$ (via nonparametric bootstrap or influence functions using the sandwich estimator \citep[see e.g.,][]{tsiatis2007semiparametric}). 

In order to apply standard group sequential methods (e.g., group sequential boundaries based on the error spending function defined by \cite{gordon1983discrete}), the covariance matrix $\boldsymbol{\Sigma}$  would need to have the independent increments structure. 
 That is, each diagonal element of $\boldsymbol{\Sigma}$ is equal to 1 and the $(k, k')$th element of $\boldsymbol{\Sigma}$, where $k'\leq k$, is equal to $\sqrt{\mathcal{I}_{k'}^*/\mathcal{I}_{k}^*}$ \citep[see e.g.,][]{scharfstein1997semiparametric, jennison1997group, jennisonturnbull1999group}. 
 The independent increments property can be equivalently formulated in terms of the  asymptotic distribution of the estimator sequence itself, i.e.,    $\widehat\theta_{t_k}$ being  asymptotically independent of all  previous increments  $\widehat\theta_{t_k}-\widehat\theta_{t_{k'}}$ for all $k'<k$, after each is centered and multiplied by $\sqrt{n}$.
 Unfortunately, an arbitrary sequence of RAL estimators $(\widehat\theta_{t_1}, \dots, \widehat\theta_{t_K})$ evaluated at analysis times $t_1<\dots<t_K$ may fail to have the independent increments property. At an intuitive level, the property may fail when estimators at different analysis times use data from the same patients.
 This was known by \cite{scharfstein1997semiparametric} and \cite{jennison1997group}, who restricted their theorems to RAL estimators that are semiparametric efficient (in which case the independent increments property is guaranteed to hold).
 
 We give covariate adjusted estimators in Appendices \ref{sec:covAdj} and \ref{sec:tmle} of the Supplementary Materials that are a generalization of the covariate adjusted estimator in \cite{ge2011covariate}, which was presented in the recent FDA draft guidance on  covariate adjustment \citep{FDA2021}. These  estimators rely on working models which are likely to be misspecified in practice, leading to estimators for which the independent increments property will generally fail to hold. This lack of independent increments can generally  occur  when estimators use working models; see e.g., \cite{rosenblum2015} for augmented inverse probability weighted estimators and \cite{shoben2014violations} for estimators based on generalized estimating equations. A long list of further examples is provided by \cite{jennison1997group} and \cite{kim2020independent}.
 To address the above problem, we  propose a  statistical method to modify any sequence $(\widehat\theta_{t_1}, \dots, \widehat\theta_{t_K})$ of (RAL) estimators  so that it will have the independent increments property and also equal or smaller variance at each time compared to the original estimator sequence. 
 
 \subsection{Method for Orthogonalizing  Sequence of RAL Estimators}\label{subsec:gsd_impl}
 At each analysis time $t_k$, our goal is to construct a new estimator $\widetilde\theta_{t_k}$ that is a linear combination of the original estimators at analysis times $t_1,\dots,t_k$ ($\widehat\theta_{t_1},\dots,\widehat\theta_{t_k}$) and that has the following properties:  (i) the new estimator $\widetilde\theta_{t_k}$ is consistent and RAL, (ii) the variance of the new estimator is decreased or left unchanged (compared to $\widehat\theta_{t_k}$ ), and (iii) the Wald test statistics corresponding with the updated sequence of estimators $(\widetilde\theta_{t_1}, \dots, \widetilde\theta_{t_K})$ have asymptotic covariance matrix with the independent increments structure.

We first present the intuition behind our method for constructing the new estimator at analysis $k$. For any real valued vector $(\lambda^{(k)}_1,\dots,\lambda^{(k)}_{k-1})$, consider the following linear combination of estimators:  
$\widehat\theta_{t_k} - \sum_{k'=1}^{k-1} \lambda^{(k)}_{k'} (\widehat\theta_{t_k}-\widehat\theta_{t_{k'}})$.
By construction, the linear combination is a consistent, RAL estimator of $\theta$ as long as each component of the original estimator sequence is as well.
We next  minimize the linear combination's  variance over all real valued vectors  $(\lambda^{(k)}_1,\dots,\lambda^{(k)}_{k-1})$, and define our updated estimator at analysis $k$ as the corresponding minimum value. This guarantees the same or better variance than the original estimator $\widehat\theta_{t_k}$ since the linear combination reduces to the original estimator if one sets each $\lambda^{(k)}_{k'}$ ($k'<k$) to 0.
Minimizing the variance over $(\lambda^{(k)}_1,\dots,\lambda^{(k)}_{k-1})$ in the above display is equivalent to 
subtracting the orthogonal ($L_2$) projection of  $\widehat\theta_{t_k}$ on the preceding increments $\widehat\theta_{t_k}-\widehat\theta_{t_{k'}}$ (after centering); this  results in the updated estimator  being orthogonal to the increments $\widehat\theta_{t_k}-\widehat\theta_{t_{k'}}$, and so also to  $\widetilde\theta_{t_k}-\widehat\theta_{t_{k'}}$ and $\widetilde\theta_{t_k}-\widetilde\theta_{t_{k'}}$.
Then, $\widetilde\theta_{t_k}$ being orthogonal (in the limit) to $\widetilde\theta_{t_k}-\widetilde\theta_{t_{k'}}$ is the independent increments property since orthogonality and independence are the same for a multivariate normal distribution which is the limiting distribution of the updated test statistics.
The above arguments are only heuristic, but we make them rigorous in our  proofs in Appendix~\ref{app:proof} of the Supplementary Materials. 

We next present, step-by-step, the proposed method for constructing the new estimator sequence. 
At the first interim analysis ($k=1$), we define $\widetilde\theta_{t_1}=\widehat\theta_{t_1}$. The corresponding test statistic equals $\widetilde Z_1=Z_1=\frac{\widehat\theta_{t_1}-\theta_0}{\widehat{se}(\widehat\theta_{t_1})}$. At each subsequent analysis $k \geq 2$: 
\begin{enumerate}
    \item We calculate $\widehat\theta_{t_k}$ and estimate the covariance matrix of $(\widehat\theta_{t_1}, \dots, \widehat\theta_{t_k})$ based on influence functions or via the nonparametric bootstrap. 
    \item Compute  $\widehat{\boldsymbol{\lambda}}^{(k)}=\left(\widehat{\lambda}^{(k)}_1,\dots,\widehat{\lambda}^{(k)}_{k-1}\right)^t$, where we define 
    \begin{equation} \left(\widehat{\lambda}^{(k)}_1,\dots,\widehat{\lambda}^{(k)}_{k-1}\right)=\arg \min_{(\lambda^{(k)}_1,\dots,\lambda^{(k)}_{k-1}) \in \mathbb{R}^{k-1}} \widehat{Var}\left\{\widehat\theta_{t_k} - \sum_{k'=1}^{k-1} \lambda^{(k)}_{k'} (\widehat\theta_{t_k}-\widehat\theta_{t_{k'}})\right\}, \label{minimization_problem} \end{equation}
    where $\widehat{Var}$ is computed using an estimate of the covariance matrix of $(\widehat\theta_{t_1}, \dots, \widehat\theta_{t_k})$.
    Then  $\widehat{\boldsymbol{\lambda}}^{(k)}=\left\{\widehat{Var}\left((\widehat\theta_{t_k}-\widehat\theta_{t_1}, \dots, \widehat\theta_{t_k}-\widehat\theta_{t_{k-1}})^t\right)\right\}^{-1}\widehat{Cov}\left(\widehat\theta_{t_k}, (\widehat\theta_{t_k}-\widehat\theta_{t_1}, \dots, \widehat\theta_{t_k}-\widehat\theta_{t_{k-1}})^t\right)$,
    where $\widehat{Cov}$ is computed using an estimate of the covariance matrix of $(\widehat\theta_{t_1}, \dots, \widehat\theta_{t_k})$.
    \item Replace $\widehat\theta_{t_k}$ by $\widetilde{\theta}_{t_k}=\widehat\theta_{t_k} - \sum_{k'=1}^{k-1} \widehat{\lambda}^{(k)}_{k'} (\widehat\theta_{t_k}-\widehat\theta_{t_{k'}})$.
    \item Estimate the variance of $\widetilde{\theta}_{t_k}$ as $$\widehat{se}(\widetilde{\theta}_{t_k})^2=(-(\widehat{\boldsymbol{\lambda}}^{(k)})^t, 1)\widehat{Var}\left((\widehat\theta_{t_k}-\widehat\theta_{t_1}, \dots, \widehat\theta_{t_k}-\widehat\theta_{t_{k-1}}, \widehat\theta_{t_k})^t\right)\left(-(\widehat{\boldsymbol{\lambda}}^{(k)})^t, 1\right)^t,$$
    and its corresponding information as $\widetilde{\mathcal{I}}_{k}=(\widehat{se}(\widetilde\theta_{t_k}))^{-2}$.
    \item Calculate $\widetilde Z_k=(\widetilde\theta_{t_k} - \theta_0)/\widehat{se}(\widetilde\theta_{t_k})$.
\end{enumerate}

 \subsection{Properties of Orthogonalized Estimators}\label{subsec:asympt}
 
 The key properties of the above orthogonalized estimators $\widetilde \theta_{t_k}$ and corresponding test statistics $\widetilde Z_k$ are given below.

\begin{theorem}[Asymptotic Properties]\label{th:main}	
Consider any sequence of RAL estimators\\ 
$(\widehat\theta_{t_1}, \dots, \widehat\theta_{t_K})$ with all components  consistent for $\theta$, and for which  \eqref{variance_convergence_assumption} holds and the covariance matrix $\boldsymbol{\Sigma}$ of the corresponding test statistics can be consistently estimated.
Then the orthogonalized estimator sequence 
$(\widetilde\theta_{t_1}, \dots, \widetilde\theta_{t_K})$ is also RAL with 
covariance matrix 
having the independent increments property. 
In addition, $\widetilde\theta_{t_k}$ at each analysis time $t_k$ is a consistent estimator for $\theta$ and has asymptotic variance less or equal to that of the original estimator $\widehat\theta_{t_k}$.
Furthermore, the analog of \eqref{variance_convergence_assumption} holds for the orthogonalized estimator sequence, i.e., $\lim_{n \rightarrow \infty}\widetilde{\mathcal{I}}_{k}/n=\lim_{n \rightarrow \infty}\left\{n\widehat{Var}(\widetilde\theta_{t_k})\right\}^{-1}=\lim_{n \rightarrow \infty}\left\{nVar(\widetilde\theta_{t_k})\right\}^{-1}=\widetilde{\mathcal{I}}^*_{k},$ with $\widetilde{\mathcal{I}}^*_{k}$ defined as the inverse of the asymptotic variance of $\widetilde\theta_{t_k}$.
The orthogonalization approach moreover ensures monotonicity of the asymptotic information $\widetilde{\mathcal{I}}^*_{k}$ and the finite sample information $\widetilde{\mathcal{I}}_{k}$ at the analysis times $t_1, \dots, t_K$, that is, $\widetilde{\mathcal{I}}^*_{k}$ and $\widetilde{\mathcal{I}}_{k}$ are non-decreasing over analysis times $t_1, \dots, t_K$.
\end{theorem}

The results from Theorem \ref{th:main} enable one to directly apply standard group sequential stopping boundaries  to the test statistics $\widetilde{Z}_k$, as long as a consistent, RAL estimator is used at each analysis (which is almost always the case for the primary efficacy analysis in confirmatory randomized trials). These stopping boundaries,
which may include (binding or non-binding) futility boundaries as well as efficacy boundaries, can be 
computed as in Section 4.2 of \cite{scharfstein1997semiparametric} and Appendix~\ref{app:bound} of the Supplementary Materials here. For example, one could apply the commonly used group sequential boundaries of  \cite{o1979multiple} or  \cite{pocock1977group}, or one could construct boundaries using  any 
 error spending function  \citep{gordon1983discrete}.
Theorem~\ref{th:main} implies that the resulting group sequential  testing procedure controls familywise Type I error rate at the desired level $\alpha$ (asymptotically). 
 Although we are working under a similar framework as \cite{scharfstein1997semiparametric} and \cite{jennison1997group}, what's new here is that we do not need the assumption that   estimators are semiparametric efficient; this enables the use of many covariate adjusted estimators within the commonly used group sequential design framework. 

The estimators at different analysis times could be chosen to be of the same type; e.g., at each analysis  an estimator that adjusts for a prespecified list of baseline variables could be used, with the only difference being that  more data are available at later analysis times.  
However, the theorem above does not require that the estimators at different time points are of the same type (though they do need to be consistent for the same estimand). E.g.,  an unadjusted estimator could be used early in the trial and a covariate adjusted estimator used later, or covariate adjusted estimators could be used throughout but adjusting for larger sets of variables at later analysis times;  
 this setup may be useful since the number of covariates that one can adjust for grows with sample size.
 In all cases, the estimators at each analysis time need to be prespecified. 

\section{Information Adaptive Design}\label{sec:inf}
A crucial question at the design stage of a clinical trial is `how much data should we gather to perform the hypothesis test at significance level $\alpha$ with power $1-\beta$?' 
The total number of participants needed to detect a clinically important treatment effect with sufficient precision often depends on nuisance parameters (e.g., probability of response in the control group for a binary endpoint) which are typically unknown before the trial starts. 
Incorrect guesses of these nuisance parameters may lead to over- or underpowered trials. 

Determining the required sample size when covariate adjusted estimators are used can be done in two ways: either a conservative assumption of no precision gain from covariate adjustment can be made (in which case any actual precision gains would increase power), or a 
projection of how much precision will be gained can be factored into the sample size calculation 
\citep{li2021estimating}. To fully leverage the precision gain resulting from covariate adjustment, however, it would be ideal to start by applying the former method for planning purposes, and then automatically adapt the sample size or duration of the trial  based on continuous monitoring of the   actual precision gain (which is directly reflected in the estimated information $\widehat{\mathcal{I}}_t$). 
  This is what we evaluate in our simulation studies and recommend for use in practice.

  Below we define information adaptive designs, which involve continuous monitoring of the  estimated information $\widehat{\mathcal{I}}_t$ to determine when to conduct interim analyses (where a trial may be stopped early for efficacy or futility) and the final analysis. Such designs can be used with unadjusted or covariate adjusted estimators (as demonstrated in our simulation studies below). 
  Our main motivation for considering information adaptive designs, however, is to apply them with covariate adjusted estimators that have been orthogonalized as in Section~\ref{sec:GSD}; 
  the combination of these approaches can lead to designs that take full advantage of precision gains from covariate adjustment (converting the gains into sample size reductions while controlling Type I error and providing the desired power).
  In other contexts, information adaptive designs and/or the key ideas underpinning them have been proposed by e.g.,  \cite{scharfstein1997semiparametric,mehta2001flexible, tsiatis2006information, Zhang2009}.

\subsection{Implementation of Information Adaptive Design}\label{subsec:inf_impl}
At the design stage, we need to specify the operating characteristics of the study such as the significance level $\alpha$, the alternative of interest $\theta_A$, along with the power $1-\beta$ to detect this alternative, and the number of interim analyses $K$ to be performed. We also need to specify a method to compute  the stopping boundaries $(c_1, \dots, c_K)$; we suggest using an error spending function  \citep{gordon1983discrete} due to its flexibility. 

After specifying the above quantities, we compute the maximum/total information needed to achieve these Type I error and power goals. For a trial without interim analyses, in order for a two-sided level-$\alpha$ test to have power  $1-\beta$ to detect the clinically important alternative $\theta_A$, we need
$\mathcal{I}(\theta_A)=\left\{(z_{\alpha/2}+z_\beta)/(\theta_A-\theta_0)\right\}^2,$
where $\mathcal{I}(\theta_A)$ denotes the required information and $z_q$ is the quantile function for the standard normal distribution. 
For a given estimator  $\widehat\theta$, its corresponding information (i.e., the reciprocal of the variance) can be estimated by $(\widehat{se}(\widehat\theta))^{-2}$. A strategy to achieve the desired power is to monitor the accrued information, estimated as $(\widehat{se}(\widehat\theta_t))^{-2}$, through time $t$ and conduct the final analysis at time $t^*$ when
$(\widehat{se}(\widehat\theta_{t^*}))^{-2}\geq\left(\frac{z_{\alpha/2}+z_\beta}{\theta_A-\theta_0}\right)^2.$
This defines the information adaptive design for trials that don't involve interim analyses.

When the data are sequentially monitored and analyzed with the possibility of early stopping, one needs to compensate for the possible loss in power resulting from accounting for  multiple testing (which is baked into the aforementioned approaches for computing stopping boundaries). The final analysis should then be conducted when the (statistical) information is equal to
$\left(\frac{z_{\alpha/2}+z_\beta}{\theta_A-\theta_0}\right)^2IF$
for a two-sided test,
where $IF>1$ denotes an inflation factor determined as a function of $K$, $\alpha$, $\beta$ and the type of error spending function \citep{kim1987design, scharfstein1997semiparametric}. 

We next define the information adaptive design for trials with multiple analysis times (i.e., GSDs).
During the trial, we monitor the data and compute the estimated information level $(\widehat{se}(\widehat\theta_t))^{-2}$ at time $t$ using all accumulated data. 
We conduct the $k$th analysis at time $t_k$ defined as the  first time that the information crosses a pre-specified information  threshold.  This defines the information adaptive design. E.g., one could use information thresholds  defined to be the following for the $k$th analysis:  $\frac{k}{K}\times\left(\frac{z_{\alpha/2}+z_\beta}{\theta_A-\theta_0}\right)^2IF$ when using two-sided alternatives and equally spaced analysis times.  


Importantly, with such an approach, we do not have to prespecify the prognostic value of the covariates nor other nuisance parameters. In addition, when the covariate adjusted estimator is more efficient (i.e., has  smaller variance) than the corresponding unadjusted estimator, covariate adjustment can lead to a shorter trial due to faster information accrual. This makes the information adaptive design as proposed by \cite{mehta2001flexible} well suited for  covariate adjusted estimators. 
The information adaptive design may  be useful beyond our context of  adjusted estimators and GSDs. Specifically, it could be useful  for unadjusted  estimators and/or  in trials without interim analyses to get correctly powered trials at the minimum possible sample size \citep{mehta2001flexible}.

In line with \cite{benkeser2020improving}, we recommend to initally set the maximum sample size for the trial conservatively, i.e., as if there would be no precision gain from covariate adjustment. We can then use the standard formulas (by positing some guesses of the other nuisance parameters) for sample size calculations in order to estimate the maximum sample size $n_{max}$ needed to achieve the Type I error and power goals. 
We suggest to use emerging data at each interim analysis time to update  $n_{max}$. This can be done periodically during the trial  using the approach of  \cite{mehta2001flexible}, who compute the new projection at analysis time $t_k$ for the maximum sample size as follows:  
$n_{max}=n(t_k)\left(\frac{z_{\alpha/2}+z_\beta}{\theta_A-\theta_0}\right)^2IF / \widehat{\mathcal{I}}_k,$
where $n(t_k)$ is the number of patients that have completed follow up at analysis time $t_k$ and $\widehat{\mathcal{I}}_k$ the corresponding information. 
Since an accurate projection of the required sample size at the final analysis should take into account the estimator to be used at that analysis, the function 
$n(t)$ and information $\widehat{\mathcal{I}}_k$ should be  calculated using the estimator that is planned for the  final analysis time (which our framework allows to differ from the estimator used at earlier times), except using the data available at time $t_k$.

Our proof in the Appendix for Theorem \ref{th:main} focuses on interim analyses at $K$ fixed time points, $t_1, \dots, t_K$. If we combine the orthogonalization approach in Section~\ref{sec:GSD} with an information adaptive design, the analysis times are not fixed since they depend on the accrued information. We require that our information adaptive algorithm selects analysis times  $\widehat{t}_1,\dots,\widehat{t}_K$ that converge to certain limit times $t^*_1,\dots,t^*_K$.
In addition, we assume that the corresponding estimator sequence at times $\widehat{t}_1,\dots,\widehat{t}_K$ has the same limit distribution 
as the corresponding estimator sequence evaluated at the limit times $t^*_1,\dots,t^*_K$.
This assumption is also implicitly made for event-driven trials and trials whose analysis times are determined by the number of patients (or primary endpoints) observed, since these are not fixed  calendar times due to their data dependence.
Discretizing time into small intervals could be used in combination with an adaptive information design that conducts each analysis at the first time that the observed information crosses a prespecified threshold; we conjecture that in this case the above assumptions will hold, as long as the estimated information converges uniformly to the limit information as sample size goes to infinity.

\section{Data Analysis and Simulation Studies Based on\\ MISTIE III Trial}\label{sec:dataAnal}
\subsection{Data Analysis}
We illustrate the proposed approaches  by using data from the open-label, blinded endpoint, Phase III clinical trial of minimally invasive surgery with thrombolysis in intracerebral haemorrhage evacuation (MISTIE III; \cite{hanley2019efficacy}). The goal was to assess whether minimally invasive catheter evacuation followed by thrombolysis, with the aim of decreasing clot size to 15 mL or less, would improve functional outcome in patients with intracerebral haemorrhage (a severe form of stroke). The primary outcome was defined as having a modified Rankin Scale (mRS) score of 0-3 measured 365 days from enrollment  (defined as a ``success"). 
Though the trial used covariate adaptive randomization, we ignore that in our discussion below, for simplicity (since analogous computations taking this into account give similar results), and we use simple randomization in our simulation study.

The estimand in the trial was defined as the (absolute) risk difference, that is, the difference between the population proportion of successes under assignment to treatment versus control (where control was standard of care using medical management).
The total sample size of approximately 500 patients was calculated based on the assumption that 25\% of the patients would have an mRS score of 0–3 in the standard medical care group versus 38\% of patients in the MISTIE group and provides a power of 88\% to detect such a population risk difference of 13\% at a 5\% significance level.
To this end, 506 patients were randomized to the MISTIE group (255) or standard medical care group (251). The analysis of the primary outcome was done in the modified intention-to-treat (mITT) population (250 in MISTIE group and 249 standard medical care group), which included all eligible, randomized patients who  were exposed to treatment.

For the mITT analysis set, adjusting for baseline variables using the targeted maximum likelihood estimator of  \cite{van2012targeted} to adjust for censoring for the primary efficacy outcome only, resulted in an absolute risk difference estimate of 0.04 ([95\% CI –0.04 to 0.12]; p=0.33). The unadjusted estimator was similar except for having a wider confidence interval.

We can now look back at the design goals of the MISTIE III trial and do a rough calculation of what an information-adaptive design would have done.
Substituting the design parameters into the equation for the maximum information for a trial without interim analyses yields
$\mathcal{I}(\theta_A)=\left(\frac{z_{0.025}+z_{0.12}}{0.13}\right)^2=582.$
Thus, the monitoring strategy described in Section~\ref{sec:inf} would call for accruing participants into the study until the total information equals or exceeds 582. 
As the standard error at the final analysis of the MISTIE III trial equals approximately 0.04, this analysis was conducted when the (estimated) information was approximately 625.
Thus, in an information-adaptive design, the final analysis would have been conducted earlier. 
In what follows, we conduct a simulation study that mimics some features of the data generating distributions from the MISTIE III trial, in order to assess the performance of the approaches proposed in Sections \ref{sec:GSD} and \ref{sec:inf}.

\subsection{Simulation Design Based on the MISTIE III Trial}\label{sec:sim}
\label{subsubsec:sim_design}

We  present our simulation study design that we use to evaluate the finite-sample properties of combining covariate adjustment 
with an information adaptive design compared to a fixed/maximum sample size trial, with and without interim analyses, for the MISTIE III  trial. We present the simulation study below using the framework/template recommended by \cite{morris2019using}.

\hspace{0.6cm}\textbf{Aims}:
To examine whether the information adaptive design in Section~\ref{sec:inf} controls the Type I error and achieves the desired power of a trial (a) without and (b) with interim analyses. In Aim (a) we also compare  the operating characteristics of the proposed trial design (which is also known as a maximum information design as there are no interim analyses) with those of maximum sample size trials. In Aim (b) we especially want to examine  whether the approach to combine covariate adjusted estimators with GSDs as explained in Section~\ref{sec:GSD} controls  (asymptotically) the Type I error and maintains the power of a trial; the timing of the analyses is determined by monitoring the information as explained in Section~\ref{sec:inf}. We moreover want to compare the operating characteristics of a GSD with timing of the analyses based on  a predefined number of primary endpoints observed (i.e., max. sample size design) versus  a group sequential design with timing of the analyses based on the observed information reaching predefined thresholds (i.e., information adaptive design).

\textbf{Data-Generating Mechanisms}: 
We construct data generating mechanisms based on resampling from the participants in the MISTIE III trial who had the primary outcome measured in order to mimic the prognostic values of baseline variables $W$ for the final outcome $Y$ (i.e., functional outcome measured at 365 days), that is, the relationships between baseline variables and outcomes observed in this trial. The baseline variables $W$ are stability intracerebral haemorrhage clot size (continuous in mL), age at randomization (continuous in years), severity of impairment at randomization as measured by GCS  (categorical with levels: 3-8, severe; 9-12, moderate; 13-15, mild), stability intraventricular haemorrhage size (continuous in mL) and intracerebral haemorrhage clot location (binary with levels lobar and deep). 
In addition, two short-term measurements on the Glasgow Rankin Scale score (categorical with levels from 0 to 6) were taken after 30 and 180 days. These are denoted by respectively $X_{30}$ and $X_{180}$.

As in the original MISTIE III trial, interest lies in testing the null hypothesis $H_0: \theta = 0$ with $\theta$ defined as
$\theta=E\left(Y|A=1\right)-E\left(Y|A=0\right)$,
at significance level $5\%$ (using a two-sided test)   with a power of 88\% under the alternative  $\theta_A=0.13$.

We consider two scenarios; in the first one there is a zero average treatment effect (i.e., the null hypothesis), while in the second one there is a positive average treatment effect (i.e., the alternative hypothesis). 
For scenario 1, we resampled data vectors $(W, X_{30}, X_{180}, Y)$ with replacement from the MISTIE III trial data. We then generated the treatment indicators $A$ independent of the data vectors $(W, X_{30}, X_{180}, Y)$ by an independent Bernoulli draw with probability 1/2 of being assigned to treatment or control. This results in an average treatment effect $\theta$ of zero as $P(Y=1|A=1)=P(Y=1|A=0)=0.43$. 
For scenario 2, we construct data generating distributions with an average treatment effect $\theta$ of 0.13, which equals the average treatment effect the MISTIE III trial was powered for. To this end, we first generated initial data vectors $(A, W, X_{30}, X_{180}, Y)$ as in scenario 1. Then, for each simulated participant with initial values $A=1$, $Y=0$ and $X_{180}=6$ (death), we randomly replaced $Y$ by an independent Bernoulli draw with probability 0.35 of being 1. 

As described above, for a trial without interim analyses (Aim (a)), the  maximum information requirement is 582.
The maximum information for trials with one interim analysis at information fraction 0.50 (Aim (b)) equals $582\cdot 1.1136 = 648$, where the inflation factor was calculated for a GSD with 1 interim analysis at information fraction 0.50 with the \texttt{R} package \texttt{rpact}. The  efficacy stopping boundaries are based on an error spending function that approximates Pocock boundaries. Here, we don't use futility boundaries. Nevertheless, in practice we recommend to use non-binding futility boundaries which are typically preferred by regulators \citep{FDA2019}.

The corresponding maximum sample size depends on an assumption for the probability of a successful outcome in the control arm as well as an assumption for the prognostic value of $W$ if we use the covariate adjusted estimators. We set the sample size as if there were no precision gain from covariate adjustment ($W$ is not prognostic). For Aim (a) two different assumptions for the probability of a successful outcome in the control arm are considered: $0.25$ (corresponding with the design assumption in the MISTIE III Trial) and $0.43$ (the assumed probability in the simulations, which is  closer to what was observed in the actual trial), which correspond with a sample sizes of respectively 498 and 578.
For Aim (b), we assume that the probability in the control arm equals $0.25$. The standard group sequential design with the analysis times based on the number of primary endpoints observed (i.e., max. sample size design) requires a total sample number of $554$, and the interim analysis is conducted when $277$ patients have their primary endpoint observed.

The uniform recruitment rate of approximately 12 patients per month corresponds with the average recruitment rate in the MISTIE III trial.
For the data generating mechanisms  above, we perform respectively $10,000$ and $100,000$ Monte Carlo runs under the alternative and null hypothesis. For computational reasons, we limit the number of runs for the targeted maximum likelihood estimator (see below) under the null hypothesis to $10,000$. 

\textbf{Targets:} See the second paragraph of ``Data Generating Mechanisms" above.

\textbf{Methods of Analysis:}
Each simulated dataset is analyzed   using Wald test statistics based on the following estimators:
\begin{itemize}
	\item Unadjusted estimator of  difference in means.
	\item Standardized logistic regression   estimator  described in Appendix \ref{sec:covAdj} of the Supplementary Materials. The logistic regression models  include as  main effects the following baseline variables:  stability intracerebral haemorrhage clot size, age, stability intraventricular haemorrhage size and intracerebral haemorrhage clot location.
	\item Longitudinal targeted maximum likelihood estimator of \cite{van2012targeted} adjusted for stability intracerebral haemorrhage clot size, age, severity of impairment as measured by GCS, stability intraventricular haemorrhage size and intracerebral haemorrhage clot location as well as two short-term measurements on the Glasgow Rankin Scale score (categorical with levels from 0 to 6) measured after 30 and 180 days. Besides adjusting for chance imbalances in pre-specified baseline variables between treatment groups, this statistical method also accounts for missing outcome data by a combination of regression modelling and inverse probability of censoring weighting using generalized linear models (without model or variable selection). More details are given in Appendix~\ref{sec:tmle} of the Supplementary Materials.
\end{itemize}
For Aim (b), 5 different test statistics are evaluated as we also consider the `updated' (defined as  having been orthogonalized) versions of the covariate adjusted estimators, following the approach in Section~\ref{sec:GSD}.


For Aim (a), we consider two maximum sample size designs  (with maximum sample sizes equal to 498 and 578, respectively) as well as an information adaptive design  (with maximum information equal to 582). For the latter design, we monitor the data and adjust $n_{max}$ every time we have the outcome available for 50 additional participants. A test is only performed once we have reached the (changing) maximum total sample size $n_{max}$. 
For Aim (b) each simulated trial dataset is analyzed as a group sequential design with one interim analysis. The timing of the interim and final analysis are based on (i) the number of participants with the primary endpoint observed (for the maximum sample size design) and (ii) the information (for the information adaptive design). The monitoring for the timing of the interim analyses happens every time we have the outcome available for 10 additional participants. We update $n_{max}$ according to the formula in Section~\ref{subsec:inf_impl} at the interim analysis and  after each subsequent batch of 50 observed outcomes. We assume that once recruitment has been stopped, it is not  restarted. For computational reasons, the monitoring of the information is based on the (estimated) information corresponding with the original estimators rather than the orthogonalized ones. 

\textbf{Performance Measures:} We assess the empirical Type I error rate and power of the test of no treatment effect (w.r.t. the risk difference) on the considered outcome $Y$ as well as the average sample size, average analysis time and average information.


\subsection{Simulation Results}
For Aim (a), the empirical power and Type I error for the different tests are summarized in Table \ref{tab:data_InfMon}; columns 3 to 6 deal with the power, the average sample number, the average analysis time and the average information under the alternative $\theta=\theta_A=0.13$, and the last four columns deal with the Type I error, the average sample number, the average analysis time and the average information under the null hypothesis $\theta=\theta_0=0$.

\begin{table}[h]
	\caption{\label{tab:data_InfMon} Results for trials with no interim analyses, comparing  information adaptive vs.  maximum sample size designs, with  3 different estimators. Goal is  88\% power. 
	}
\centering
\begin{tabular}{ l  l c  c c c ccc cc}
  \hline
  && \multicolumn{9}{ c }{Simulation Parameter}\\
  && \multicolumn{4}{ c }{$\theta=0.13$ (Alternative)} && \multicolumn{4}{ c }{$\theta=0$ (Null)}\\
  \cline{3-6}  \cline{8-11} 
 Design Type&& Power & \textbf{ASN} & AAT & AI && Type I & \textbf{ASN} & AAT & AI\\
  \hline  
  Info. Adaptive & Unadj. &88.4\%&\textbf{571} &1876 &582 & & 5.28\% &\textbf{569} &1871 &582\\
  with $\mathcal{I}(\theta_A)=582$ & Stand. &87.3\%&\textbf{433} &1509 &567 & & 5.28\% &\textbf{402} &1427&568\\
   & TMLE &87.5\%&\textbf{432} &1506 &571 & & 5.11\% &\textbf{402} &1428 &574\\ \\
Max. Sample Size& Unadj. &83.1\%&- &1683 &508 & & 5.14\% &- &1682 &509\\
   with $n_{max}=498$ & Stand.&91.1\%&- &1682 &652 & & 5.14\% &- &1682 &705\\
   & TMLE &91.7\%&- &1682 &659 & & 4.92\% &- &1682&713\\ \\
   Max. Sample Size & Unadj. &88.4\%&- &1894 &589 & & 5.24\% &- &1894&591\\
   with $n_{max}=578$ & Stand. &94.5\%&- &1893 &759 & & 5.13\% &- &1894 &821\\
   & TMLE &94.6\%&- &1893 &768 & & 5.14\% &- &1894&829\\ 
   \hline
\end{tabular}
 {\raggedright Unadj., unadjusted estimator; Stand., standardization estimator; TMLE, targeted maximum likelihood estimator; ASN, average sample number; AAT, average analysis time (in days); AI, average information. \par}
\end{table}

The results show that the information adaptive design achieves the desired power of 88\% under the alternative $\theta_A=0.13$ for all three estimators. 
For the maximum sample size design, the  power depends on whether the true probability under control is smaller, larger or equal to  its assumed value at the design stage. 
For the unadjusted estimator and maximum sample size design, it requires 578 patients to achieve 88\% power.  The covariate adjusted estimators combined with the maximum sample size design at sample size 578  achieve a higher power of respectively 94.5\% and 94.6\% for the standardization and targeted maximum likelihood estimators. This increase in power is a result of the increase in (average) information (i.e., higher precision) due to covariate adjustment (average information of 589 for the unadjusted estimator compared to 759 and 768 for the covariate adjusted estimators). 
The maximum sample size design with 498 patients is not sufficient for the unadjusted estimator to achieve the 88\% power as desired; only 83.1\% of the 10,000 simulations are able to reject the null hypothesis. This underpower is a consequence of assuming a too low probability under control at the design stage, resulting in a too small sample size and information. 
The simulation results under the null hypothesis ($\theta=0$) show that the Type 1 error of  information adaptive and maximum sample size  designs are (approximately) correct.

The information adaptive design allows one to fully employ efficiency gains from covariate adjustment, which here led to a shorter average duration and average sample number while maintaining the desired Type I error and power. On average, we observe a 24\% and 29\% reduction in sample size due to covariate adjustment under the alternative and null hypothesis, respectively. On the other hand, when conducting a maximum sample size design, efficiency gains due to covariate adjustment do not impact sample size or trial duration, but instead translate into higher power compared to the unadjusted estimator; the downside is that under the null hypothesis, the efficiency gains due to covariate adjustment are essentially wasted, unlike for the information adaptive design. 


\begin{table}[h!]
	\caption{\label{tab:data_sim_gsd} 
	Results for GSD's, comparing  information adaptive vs.  maximum sample size designs, with  3 different estimators. Goal is  88\% power. 
	}
\centering
\begin{tabular}{ l  l   c c c c  c cc}
  \hline
  Design Type&& Power  & \textbf{ASN} & ASN1 & ASN2 & AAT& AAT1 & AAT2   \\
  \hline
  && \multicolumn{7}{ c }{$\theta=0.13$ (Alternative)}\\
  \cline{3-9} 
  Max. Sample Size & Unadj. & 83.6\% &\textbf{485}  &415  &554   & 1465 &1098  &1831   \\
    with $\widehat\theta_{t_k}$& Stand. &91.1\% & \textbf{466} &  415& 554  & 1364&  1098&  1830\\
    & TMLE &91.5\%   & \textbf{453}& 415 & 554 & 1297 & 1098 & 1830\\ 
    Max. Sample Size & Stand. & 91.1\% &  \textbf{466}& 415 & 554  & 1364&  1098&  1830\\
    with $\widetilde\theta_{t_k}$& TMLE &91.3\%   & \textbf{453}& 415 & 554 & 1297 & 1098 & 1830\\ \\
   Information Adaptive & Unadj. & 88.3\% & \textbf{534} & 460 & 636  & 1566& 1218 & 2047 \\
    with $\widehat\theta_{t_k}$& Stand. &87.3\% & \textbf{428} & 387 & 483  & 1284& 1024 & 1641 \\
    & TMLE & 88.1\% & \textbf{403} & 347 & 484  & 1214& 917 & 1644 \\ 
    Information Adaptive  & Stand. &87.2\% & \textbf{428} & 387 & 483  & 1284& 1024 & 1641 \\
    with $\widetilde\theta_{t_k}$& TMLE & 88.0\% & \textbf{403} & 347 & 484  & 1214& 917 & 1644  \\ \\
      && \multicolumn{7}{ c }{$\theta=0$ (Null)}\\
  \cline{3-9} 
    Max. Sample Size & Unadj. & 5.18\% & \textbf{550} & 415 & 554  & 1809 & 1098 & 1831 \\
    with $\widehat\theta_{t_k}$& Stand. &5.54\% & \textbf{549} & 415  & 554  & 1803& 1098 & 1831  \\
    & TMLE &5.72\%   &\textbf{548} & 415 & 554 &  1799& 1098 & 1830\\
    Max. Sample Size & Stand. & 5.54\% & \textbf{549} & 415  & 554  & 1803&  1098&  1831\\
    with $\widetilde\theta_{t_k}$& TMLE &5.66\%   &\textbf{548} & 415 & 554 &  1799& 1098 & 1830\\ \\
    Information Adaptive & Unadj. & 5.29\% & \textbf{628} & 459 & 634 & 2014& 1215 & 2042 \\
    with $\widehat\theta_{t_k}$& Stand. & 5.42\%  & \textbf{446} & 370 & 449  & 1532& 978 & 1552\\
    & TMLE & 5.26\%  & \textbf{445} & 328 & 449 & 1528& 867 & 1552 \\
    Information Adaptive & Stand. & 5.41\%  & \textbf{446} & 370 & 449  & 1532& 978 & 1552 \\
    with $\widetilde\theta_{t_k}$& TMLE & 5.24\% & \textbf{445} & 328 & 449   & 1528& 867 & 1552\\\\ 
\hline
\end{tabular}
{\raggedright Unadj., unadjusted estimator; Stand., standardization estimator; TMLE, targeted maximum likelihood estimator; ASN, average sample number; ASNj, average sample number at analysis $j$ ($j=1,2$); AAT, average analysis time (in days); AATj, average analysis time (in days) of analysis $j$ ($j=1,2$). \par}
\end{table}

For Aim (b), the empirical power, the Type I error, the average sample numbers and the average analyses times for the different estimators are summarized in Table \ref{tab:data_sim_gsd}. 
For the information adaptive design, the average analysis time and sample number are considerably reduced for the covariate adjusted estimators compared to the unadjusted estimators, at both analyses. Specifically, for the standardization estimator, we observe an average reduction of 20\% and 29\% in sample size due to covariate adjustment under the alternative and null hypothesis, respectively. When using the targeted maximum likelihood estimator, these reductions are approximately 25\% and 29\%. The (small) differences between the two covariate adjusted estimators are a consequence of the fact that the targeted maximum likelihood estimator also incorporates information on intermediate outcomes and an additional baseline covariate.

The results show that the desired power of 88\% is achieved for the information adaptive group sequential design in combination with the different covariate adjusted estimators as well as the unadjusted estimator.
For a max. sample size group sequential design (i.e., with analysis times based on number of primary endpoints observed), the precision gain due to covariate adjustment leads to a higher power compared to the unadjusted estimator. As before, the actual power depends on the sample size which relies on assumptions for the probability  of a successful  outcome under control and the prognostic value of the baseline covariates (not shown here).
In addition, the Type I error is approximately controlled regardless of the estimator. 

The small inflation in Type I error seems to decrease in larger samples (Appendix~\ref{app:addSim_large} of the Supplementary Materials) or when a small sample correction is used (Appendix~\ref{app:addSim_correction} of the Supplementary Materials). This inflation is not a result of a violation of the independent increments. On the contrary, this assumption appeared (surprisingly) to be satisfied in these simulations as the correlation between $\widehat{\theta}_2$ and $\widehat{\theta}_2-\widehat{\theta}_1$ were not significantly different from zero (at the 1\% significance level). This is also the reason why no real improvement from the orthogonalization is observed. 
Additional simulation results show the improvement when the independent increments property is not satisfied for the original estimators (Appendix~\ref{app:sec:sim} of the Supplementary Materials). Finally, in Appendix~\ref{app:addSim_prognostic} of the Supplementary Materials we demonstrate how well the information adaptive design is able to perform for different prognostic values in the baseline variables.

\section{Discussion}\label{sec:disc}

Our proposal to combine covariate adjustment with group sequential, information adaptive designs can help to  achieve faster, more efficient trials  without sacrificing validity or power. In particular, our approach can lead to faster trials even when the experimental treatment is ineffective; this may be more ethical in settings where it is desirable to stop as early as possible to avoid unnecessary exposure to side effects.

An alternative approach to apply group sequential methods in clinical trials, is to use the estimated covariance matrix to compute group sequential boundaries based on a multivariate integration \citep{kim2020independent}. Although the MULNOR program \citep{schervish1984multivariate} can be employed for this, such an approach has the drawback of being computationally intensive. 
The multivariate integration approach also has the drawback of being more challenging for trial planning. For example, it cannot be directly combined with an information adaptive design as it is challenging to determine the inflation factor to account for the interim analyses. Finally, it lacks the simplicity of our proposal as it doesn't allow application of standard group sequential methods such as the \cite{pocock1977group} and \cite{o1979multiple} stopping boundaries or the \cite{gordon1983discrete} error spending function. 

Although the simulation study has focused on estimators for binary endpoints, the approach can be used for all kind of endpoints (e.g., continous, ordinal, time-to-event, \dots) and estimands as long as the considered estimators are consistent RAL estimators (see Section~\ref{sec:notation}). In particular, our method can be applied to targeted maximum likelihood estimators.
Moreover, all of the theoretical results in this paper can be extended to handle combined use of stratified randomization (which is commonly done in practice) and covariate adjustment by using the general technique from \cite{wang2021model}.
In addition, our approach can be expanded to handle missing data due to drop-out under the missing at random assumption (conditional on the covariates and treatment assignment) by using doubly robust methods \citep[see e.g., Appendix B in][]{benkeser2020improving}.

Based on our simulation studies, we don't have an example where Type I error is highly inflated without using our proposal in Section~\ref{sec:GSD}. 
Nevertheless, as we do not know the underlying data-generating mechanism and as the problem may potentially be worse for other data-generating mechanisms, it will be difficult to know in advance how big the Type I error inflation will be without our method. It is therefore safer to use the proposal as it guarantees to maintain the Type I error in large samples.

In the simulation studies, we saw a small inflation in the Type I error for our proposal (as well as other methods). This might be a consequence of on the one hand the unblinded information monitoring \citep[see e.g.,][]{friede2012blinded} and on the other hand the erratic behavior of standard Wald tests for a binomial proportion \citep[see e.g.,][]{brown2001interval}. 
However, additional simulation studies (see Appendix~\ref{app:addSim_correction} of the Supplementary Materials) show that the variance estimator proposed in \cite{tsiatis2008covariate} to correct for the estimation of nuisance parameters in small samples decreases this inflation. 
As the nonparametric BCa bootstrap \citep{efron1994introduction} has shown to improve the results for covariate adjusted estimators \citep{benkeser2020improving}, this is another direction for future research. 
We also want to investigate the performance of the proposed methods for permutation based inference as they provide exact control of the Type I error.
In addition, blinded continuous information  monitoring that does not require breaking the treatment code may help to decrease the  inflation in Type I error in smaller samples \citep{friede2012blinded}. Blinded estimation for the standardization estimator in the simulations (see Appendix~\ref{sec:covAdj} of the Supplementary Materials) is discussed in \cite{VanLancker2020}.

We considered covariate adjusted estimators that use 
prespecified baseline variables. 
Another option is to  use a prespecified  variable selection  algorithm (i.e.,  a data adaptive method). A future research direction is to investigate how the proposed methods can be combined with variable and/or model selection. 

\texttt{R} functions to implement the methodology described in Section~\ref{sec:GSD} and Section~\ref{sec:inf} are available on Github at (GitHub link removed so as not to reveal author names).

\appendix

\section{Derivation and Asymptotics of Minimization \\Procedure Used in Section 3.1}
\subsection{Derivation of  Variance Minimizer Formula Used in \\Section~\ref{subsec:gsd_impl}}\label{app:lambdahat}
In Section~\ref{subsec:gsd_impl}, we presented the minimizer of the quantity  
\begin{align*}
    \widehat{Var}\{\widehat\theta_{t_k} - \sum_{k'=1}^{k-1} \lambda^{(k)}_{k'} (\widehat\theta_{t_k}-\widehat\theta_{t_{k'}})\}&=\widehat{Var}\{\widehat\theta_{t_k}\}+(\boldsymbol{\lambda}^{(k)})^t\widehat{Var}\{(\widehat\theta_{t_k}-\widehat\theta_{t_1}, \dots, \widehat\theta_{t_k}-\widehat\theta_{t_{k-1}})^t\}\boldsymbol{\lambda}^{(k)}\\
    &-2(\boldsymbol{\lambda}^{(k)})^t\widehat{Cov}\{\widehat\theta_{t_k}, (\widehat\theta_{t_k}-\widehat\theta_{t_1}, \dots, \widehat\theta_{t_k}-\widehat\theta_{t_{k-1}})^t\},
\end{align*}
with respect to $\boldsymbol{\lambda}^{(k)}=\left(\lambda^{(k)}_1,\dots,\lambda^{(k)}_{k-1}\right)^t$. We now give the derivation of our formula (in Section~\ref{subsec:gsd_impl}) for this minimizer. 

First, we find the derivative (in denominator layout) with respect to $\boldsymbol{\lambda}^{(k)}$
\begin{align*}
    \frac{\partial}{\partial \boldsymbol{\lambda}^{(k)}}\widehat{Var}\{\widehat\theta_{t_k} - \sum_{k'=1}^{k-1} \lambda^{(k)}_{k'} (\widehat\theta_{t_k}-\widehat\theta_{t_{k'}})\}&=\frac{\partial}{\partial \boldsymbol{\lambda}^{(k)}}\widehat{Var}\{\widehat\theta_{t_k}\}\\
    &+\frac{\partial}{\partial \boldsymbol{\lambda}^{(k)}}(\boldsymbol{\lambda}^{(k)})^t\widehat{Var}\{(\widehat\theta_{t_k}-\widehat\theta_{t_1}, \dots, \widehat\theta_{t_k}-\widehat\theta_{t_{k-1}})^t\}\boldsymbol{\lambda}^{(k)}\\
    &-2\frac{\partial}{\partial \boldsymbol{\lambda}^{(k)}}(\boldsymbol{\lambda}^{(k)})^t\widehat{Cov}\{\widehat\theta_{t_k}, (\widehat\theta_{t_k}-\widehat\theta_{t_1}, \dots, \widehat\theta_{t_k}-\widehat\theta_{t_{k-1}})^t\}\\
    &=0+2\widehat{Var}\{(\widehat\theta_{t_k}-\widehat\theta_{t_1}, \dots, \widehat\theta_{t_k}-\widehat\theta_{t_{k-1}})^t\}\boldsymbol{\lambda}^{(k)}\\
    &-2\widehat{Cov}\{\widehat\theta_{t_k}, (\widehat\theta_{t_k}-\widehat\theta_{t_1}, \dots, \widehat\theta_{t_k}-\widehat\theta_{t_{k-1}})^t\}.
\end{align*}
We then set it to zero at the optimum $\widehat{\boldsymbol{\lambda}}^{(k)}=\left(\widehat{\lambda}^{(k)}_1,\dots,\widehat{\lambda}^{(k)}_{k-1}\right)^t$,
$$2\widehat{Var}\{(\widehat\theta_{t_k}-\widehat\theta_{t_1}, \dots, \widehat\theta_{t_k}-\widehat\theta_{t_{k-1}})^t\}\widehat{\boldsymbol{\lambda}}^{(k)}-2\widehat{Cov}\{\widehat\theta_{t_k}, (\widehat\theta_{t_k}-\widehat\theta_{t_1}, \dots, \widehat\theta_{t_k}-\widehat\theta_{t_{k-1}})^t\}=\boldsymbol{0}.$$
Solving this, we get
$\widehat{\boldsymbol{\lambda}}^{(k)}=\left\{\widehat{Var}\left((\widehat\theta_{t_k}-\widehat\theta_{t_1}, \dots, \widehat\theta_{t_k}-\widehat\theta_{t_{k-1}})^t\right)\right\}^{-1}\widehat{Cov}\left(\widehat\theta_{t_k}, (\widehat\theta_{t_k}-\widehat\theta_{t_1}, \dots, \widehat\theta_{t_k}-\widehat\theta_{t_{k-1}})^t\right)$.

\subsection{(Heuristic) Overview of  Asymptotics of Minimization\\ Procedure}\label{app:lambda*}
The aim of the minimization procedure in Step 2 of the algorithm in Section~\ref{subsec:gsd_impl} of the main article is to find at each analysis $k \geq 2$  the values $(\lambda^{(k)}_1,\dots,\lambda^{(k)}_{k-1}) \in \mathbb{R}^{k-1}$ that equal
$$\arg \min_{(\lambda^{(k)}_1,\dots,\lambda^{(k)}_{k-1}) \in \mathbb{R}^{k-1}} Var\{\widehat\theta_{t_k} - \sum_{k'=1}^{k-1} \lambda^{(k)}_{k'} (\widehat\theta_{t_k}-\widehat\theta_{t_{k'}})\},$$
which is equivalent to
$$\arg \min_{(\lambda^{(k)}_1,\dots,\lambda^{(k)}_{k-1}) \in \mathbb{R}^{k-1}} \widehat{\mathcal{I}}_kVar\{\widehat\theta_{t_k} - \sum_{k'=1}^{k-1} \lambda^{(k)}_{k'} (\widehat\theta_{t_k}-\widehat\theta_{t_{k'}})\}$$
as multiplying the variance with the information $\widehat{\mathcal{I}}_k$ will not impact the minimizer. This can then be approximated by
$$\arg \min_{(\lambda^{(k)}_1,\dots,\lambda^{(k)}_{k-1}) \in \mathbb{R}^{k-1}} Var\left[\widehat{\mathcal{I}}_k^{1/2}(\widehat\theta_{t_k}-\theta_0) - \sum_{k'=1}^{k-1} \lambda^{(k)}_{k'} \{\widehat{\mathcal{I}}_k^{1/2}(\widehat\theta_{t_k}-\theta_0)-\widehat{\mathcal{I}}_k^{1/2}(\widehat\theta_{t_{k'}}-\theta_0)\}\right],$$
which can be rewritten as
$$\arg \min_{(\lambda^{(k)}_1,\dots,\lambda^{(k)}_{k-1}) \in \mathbb{R}^{k-1}} Var\{Z_k - \sum_{k'=1}^{k-1} \lambda^{(k)}_{k'} (Z_k-(\widehat{\mathcal{I}}_k/\widehat{\mathcal{I}}_{k'})^{1/2}Z_{k'})\},$$
by the definition of $Z_k$.

The assumptions in the main paper imply  that $\left(Z_1, \dots, Z_K\right)$ converges in distribution to $\left(Z^*_1, \dots, Z^*_K\right)$, which is multivariate normal distributed with mean vector $\boldsymbol{\delta}$ and covariance matrix $\boldsymbol\Sigma$.
We moreover assume that $\lim_{n \rightarrow \infty}\widehat{\mathcal{I}}_{k}/n=\mathcal{I}_{k}^*$, the vector $(\widehat{\mathcal{I}}_1/\widehat{\mathcal{I}}_K,\dots,\widehat{\mathcal{I}}_{K}/\widehat{\mathcal{I}}_K)$ converges to the constant vector $(f_1, \dots, f_{K})=(\mathcal{I}_1^*/\mathcal{I}_K^*,\dots,\mathcal{I}_{K}^*/\mathcal{I}_K^*)$. 
Then, roughly speaking, asymptotically, the minimization problem   reduces to finding $\left(\lambda^{*(k)}_1,\dots,\lambda^{*(k)}_{k-1}\right)$ for which
\begin{eqnarray}
\left(\lambda^{*(k)}_1,\dots,\lambda^{*(k)}_{k-1}\right)
& = &\arg \min_{(\lambda^{(k)}_1,\dots,\lambda^{(k)}_{k-1}) \in \mathbb{R}^{k-1}} Var[Z^*_k - \sum_{k'=1}^{k-1} \lambda^{(k)}_{k'} \{Z^*_k-(f_k/f_{k'})^{1/2}Z^*_{k'}\}]. \label{eq:minimizer} 
\end{eqnarray}

To compute the minimizer $\boldsymbol{\lambda}^{*(k)}=\left(\lambda^{*(k)}_1,\dots,\lambda^{*(k)}_{k-1}\right)^t$ in Equation \eqref{eq:minimizer}, observe that it has the form of a linear least squares regression problem as the regressors $Z^*_k-(f_k/f_{k'})^{1/2}Z^*_{k'}$ for all $k'<k$ have mean zero because of the consistency of $\widehat\theta_{t_{1}},\dots, \widehat\theta_{t_{K}}$. In particular, it is equivalent to regressing $Y^{(k)} = Z^*_k$ on the $k \times 1$ vector $\mathbf{X}^{(k)}=(Z^*_k-(f_k/f_{1})^{1/2}Z^*_{1},\dots,Z^*_k-(f_k/f_{k-1})^{1/2}Z^*_{k-1})^t$, resulting in the closed form solution $\boldsymbol{\lambda}^{*(k)}=\{E(\mathbf{X^{(k)}(X^{(k)})^t})\}^{-1}E(Y^{(k)}\mathbf{X}^{(k)})$. 

\section{Proof of Theorem 1}\label{app:proof}
In this Appendix we prove that for the centered (i.e., substracting the true parameter $\theta$ instead of the value under the null, $\theta_0$) test statistics $\left(\widetilde Z_1, \dots, \widetilde Z_K\right)^t$ it holds that
$$ \left(\widetilde Z_1, \dots, \widetilde Z_K\right)^t \xrightarrow{D} N\left(\boldsymbol{0}, \widetilde{\boldsymbol{\Sigma}}\right), $$
with $\widetilde{\boldsymbol{\Sigma}}$ a $K\times K$ matrix with an independent increment structure; i.e., it has 1's on its main diagonal and $\widetilde{\boldsymbol{\Sigma}}_{{k'}k} = 
(\widetilde f_{\min\{{k'},k\}}/\widetilde f_{\max\{{k'},k\}})^{1/2}$ for all $k',k \in \{1, \dots, K\}$. Here, $\widetilde f_k$ is the probability limit of $\widehat{se}(\widetilde{\theta}_{t_K})^2/\widehat{se}(\widetilde{\theta}_{t_k})^2$.

In particular, we first verify the consistency, asymptotic linearity and asymptotic normality of $\widetilde\theta_{t_k}$. We then show the aforementioned properties about independent increments and that $\widetilde\theta_{t_k}$ is asymptotically as or more precise than the original estimator $\widehat\theta_{t_k}$. Finally, we prove the monotonicity of the (asymptotic and estimated) information corresponding with the sequence of orthogonalized estimators.

\subsection*{Consistency}
Assuming that Equation (1) in the main article holds and that $\boldsymbol{\Sigma}$ can be consistently estimated, 
we can consistently estimate $E(\mathbf{X}^{(k)}(\mathbf{X}^{(k)})^t)=Var(\mathbf{X}^{(k)})$ and $E(Y^{(k)}\mathbf{X}^{(k)})=Cov(Y^{(k)},\mathbf{X}^{(k)})$.
It then follows from the continuous mapping theorem and Slutsky's theorem that $\widehat{\boldsymbol{\lambda}}^{(k)}$ converges in probability to $\boldsymbol{\lambda}^{*(k)}$.
Under the assumption that $\widehat\theta_{t_k}$ is consistent, that is, converges in probability to $\theta$ ($k\in \{1, \dots, K\}$), 
\begin{eqnarray*}
    \widehat\theta_{t_k} - \sum_{k'=1}^{k-1} \widehat\lambda^{(k)}_{{k'}} (\widehat\theta_{t_k}-\widehat\theta_{t_{k'}})\overset{p}{\to} \theta
\end{eqnarray*}
by Slutsky's theorem and the fact that $\widehat{\boldsymbol{\lambda}}^{(k)}$  converges to its probability limit (i.e., $\boldsymbol{\lambda}^{*(k)}$).

\subsection*{Asymptotic Linearity and Normality}
Assuming that $(\widehat\theta_{t_1}, \dots, \widehat\theta_{t_K})$ are asymptotic linear estimators, that is,
\begin{eqnarray*}
    \sqrt{n}(\widehat\theta_{t_k}-\theta)=\frac{1}{\sqrt{n}}\sum_{i=1}^n\phi_{t_k}(X_{t_k,i})+o_p(1),
\end{eqnarray*}
where $\phi_{t_k}(X_{t_k,i})$ is the influence function of $\widehat\theta_{t_k}$ and $X_{t_k,i}$ are the data at time $t_k$ for patient $i$. First, for simplicity, consider the case where the minimizer  $\boldsymbol{\lambda}^{(k)}=(\lambda^{(k)}_1, \dots, \lambda^{(k)}_{k-1})'$ in Equation \eqref{eq:minimizer} is known and equal to $\boldsymbol{\lambda}^{*(k)}=\left(\lambda^{*(k)}_1,\dots,\lambda^{*(k)}_{k-1}\right)'$,
\begin{eqnarray*}
    \sqrt{n}(\widetilde{\theta}_{t_k}-\theta)&=&\sqrt{n}\left(\widehat\theta_{t_k} - \sum_{k'=1}^{k-1} \lambda^{*(k)}_{k'} (\widehat\theta_{t_k}-\widehat\theta_{t_{k'}})-\theta\right)\\
    &=&\sqrt{n}\left(\widehat\theta_{t_k} -\theta - \sum_{k'=1}^{k-1} \lambda^{*(k)}_{k'} (\widehat\theta_{t_k}-\theta-(\widehat\theta_{t_{k'}}-\theta))\right)\\
    &=&\frac{1}{\sqrt{n}}\sum_{i=1}^n\left(\phi_{t_k}(X_{t_k,i})- \sum_{k'=1}^{k-1} \lambda^{*(k)}_{k'} (\phi_{t_k}(X_{t_k,i})-\phi_{t_{k'}}(X_{t_{k'},i}))\right)+o_p(1),
\end{eqnarray*}
by Slutksy's theorem. Define the corresponding influence function for $\widetilde{\theta}_{t_k}$ at 
 $\boldsymbol{\lambda}^{(k)} = \boldsymbol{\lambda}^{*(k)}$ as 
$$\widetilde{\phi}_{t_k}(X_{t_k}; \boldsymbol{\lambda}^{(k)})=\phi_{t_k}(X_{t_k})- \sum_{k'=1}^{k-1} \lambda^{(k)}_{k'} (\phi_{t_k}(X_{t_k})-\phi_{t_{k'}}(X_{t_{k'}})).$$

Next, consider the (more realistic) case where the minimizer in Equation \eqref{eq:minimizer} is unknown and is  estimated. Then, from a standard Taylor expansion (see Chapter 3 in \cite{tsiatis2007semiparametric}), the influence function  of $\widetilde\theta_{t_k}$ is 
$$
\widetilde{\phi}_{t_k}(X_{t_k}; \boldsymbol{\lambda}^{*(k)}) - E\left(\frac{\partial \widetilde{\phi}_{t_k}(X_{t_k}; \boldsymbol{\lambda}^{(k)})}{\partial \boldsymbol{\lambda}^{(k)}}\biggr\rvert_{\boldsymbol{\lambda}^{(k)}=\boldsymbol{\lambda}^{*(k)}}\right)E^{-1}\left(\frac{\partial U_{t_k}(X_{t_k}; \boldsymbol{\lambda}^{(k)})}{\partial \boldsymbol{\lambda}^{(k)}}\biggr\rvert_{\boldsymbol{\lambda}^{(k)}=\boldsymbol{\lambda}^{*(k)}}\right)U_{t_k}(X_{t_k}; \boldsymbol{\lambda}^{*(k)}),
$$
with $U_{t_k}(X_{t_k}; \boldsymbol{\lambda}^{(k)})$ the estimating equation of $\boldsymbol{\lambda}^{*(k)}$. Consequently, as $E\left(\frac{\partial \widetilde{\phi}_{t_k}(X_{t_k}; \boldsymbol{\lambda}^{(k)})}{\partial \boldsymbol{\lambda}^{(k)}}\biggr\rvert_{\boldsymbol{\lambda}^{(k)}=\boldsymbol{\lambda}^{*(k)}}\right)=\mathbf{0}$,
\begin{eqnarray*}
       \sqrt{n}(\widetilde{\theta}_{t_k}-\theta)&=&\frac{1}{\sqrt{n}}\sum_{i=1}^n\left(\phi_{t_k}(X_{t_k,i})- \sum_{k'=1}^{k-1} \lambda^{*(k)}_{k'} (\phi_{t_k}(X_{t_k,i})-\phi_{t_{k'}}(X_{t_{k'},i}))\right)+o_p(1).
\end{eqnarray*}
This proves the asymptotic linearity of the proposed estimators. 
The asymptotic normality follows from the multivariate central limit theorem and Slutsky's theorem. 

\subsection*{Independent Increments Property}
As $\lim_{n \rightarrow \infty}\widehat{\mathcal{I}}_{k}/n=\lim_{n \rightarrow \infty}\left\{n\widehat{Var}(\widehat\theta_{t_k})\right\}^{-1}=\mathcal{I}_{k}^*$, at the truth (i.e., substracting the true parameter $\theta$ instead of the value under the null, $\theta_0$) the test statistic $Z_k$ admits the expansion $\sum_{i=1}^n\left(\mathcal{I}_{k}^*\right)^{1/2}\phi_{t_k}(X_{t_k,i})/\sqrt{n}+o_p(1)$. 
From the part on asymptotic linearity, we know that $\widetilde\theta_{t_k}$ has influence function
$\widetilde{\phi}_{t_k}(X_{t_k}; \boldsymbol{\lambda}^{*(k)})=\phi_{t_k}(X_{t_k})- \sum_{k'=1}^{k-1} \lambda^{*(k)}_{k'} (\phi_{t_k}(X_{t_k})-\phi_{t_{k'}}(X_{t_{k'}}))$.
Moreover, we let $\widetilde{\mathcal{I}}_{k}=1/\widehat{se}(\widetilde{\theta}_{t_k})^2$ and define $\widetilde{\mathcal{I}}^*_{k}$ as $\lim_{n \rightarrow \infty}\left\{nVar(\widetilde\theta_{t_k})\right\}^{-1}=\lim_{n \rightarrow \infty}\left\{n\cdot(-(\boldsymbol{\lambda}^{*(k)})^t, 1)Var\left((\widehat\theta_{t_k}-\widehat\theta_{t_1}, \dots, \widehat\theta_{t_k}-\widehat\theta_{t_{k-1}}, \widehat\theta_{t_k})^t\right)(-(\boldsymbol{\lambda}^{*(k)})^t, 1)^t\right\}^{-1}$.
Assuming that Equation (1) in the main article holds, that $\boldsymbol{\Sigma}$ can be consistently estimated and consequently that $\widehat{\boldsymbol{\lambda}}^{(k)}$ is a consistent estimator for $\boldsymbol{\lambda}^{*(k)}$, it holds that $\lim_{n \rightarrow \infty}\widetilde{\mathcal{I}}_{k}/n=\lim_{n \rightarrow \infty}\left\{n\widehat{Var}(\widetilde\theta_{t_k})\right\}^{-1}=\lim_{n \rightarrow \infty}\left\{nVar(\widetilde\theta_{t_k})\right\}^{-1}=\widetilde{\mathcal{I}}^*_{k}$ by Slutsky's theorem.
Consequently,
the test statistic $\widetilde Z_k$ admits the expansion $\sum_{i=1}^n\left(\widetilde{\mathcal{I}}^*_{k}\right)^{1/2}\widetilde\phi_{t_k}(X_{t_k,i}; \boldsymbol{\lambda}^{*(k)})/\sqrt{n}+o_p(1)$, at the truth (i.e., substracting the true parameter $\theta$ instead of the value under the null, $\theta_0$).

For the independent increments property to hold, we need to prove that $\widetilde Z_k$ is asymptotically independent of $\widetilde Z_k - (\widetilde f_k/\widetilde f_{k'})^{1/2}\widetilde Z_{k'}$ for all $k'<k$. As $\widetilde f_k=\widetilde{\mathcal{I}}^*_{k}/\widetilde{\mathcal{I}}^*_{K}$, this is equivalent to proving that
\begin{align*}
    E\left\{\widetilde{\phi}_{t_k}(X_{t_k})\left(\widetilde{\phi}_{t_k}(X_{t_k})-\widetilde{\phi}_{t_{k'}}(X_{t_{k'}})\right)\right\}=0.
\end{align*}

Let $T_k=(\lambda^{*(k)}_1, \dots, \lambda^{*(k)}_{k-1}, 1)^t$, with $(\lambda^{*(k)}_1, \dots, \lambda^{*(k)}_{k-1})^t$ the minimizer of Equation \eqref{eq:minimizer}. 
From Equation \eqref{eq:minimizer}, we know that $Z_k - \sum_{k'=1}^{k-1} \lambda^{*(k)}_{k'} \{Z_k-(f_k/f_{k'})^{1/2}Z_{k'}\}$ is asymptotically orthogonal to 
$\{Z_k-(f_k/f_{k'})^{1/2}Z_{k'}\}$ for all $k'=1, \dots, k-1$; that is, $Z^*_k - \sum_{k'=1}^{k-1} \lambda^{*(k)}_{k'} \{Z^*_k-(f_k/f_{k'})^{1/2}Z^*_{k'}\}$ is independent of $\{Z^*_k-(f_k/f_{k'})^{1/2}Z^*_{k'}\}$.
As $f_k=\mathcal{I}^*_{k}/\mathcal{I}^*_{K}$, it then follows that 
$$\frac{1}{\sqrt{n}}\sum_{i=1}^n\mathcal{I}^*_{k}\left(\phi_{t_k}(X_{t_k,i})- \sum_{k'=1}^{k-1} \lambda^{*(k)}_{k'} (\phi_{t_k}(X_{t_k,i})-\phi_{t_{k'}}(X_{t_{k'},i}))\right)$$
is independent of
$$\frac{1}{\sqrt{n}}\sum_{i=1}^n\mathcal{I}^*_{k}\left(\phi_{t_k}(X_{t_k,i})-\phi_{t_{k'}}(X_{t_{k'},i})\right).$$
Thus, for each $k'=1, \dots, k-1$,
\begin{align}\label{eq:proof1}
E\left\{\widetilde{\phi}_{t_k}(X_{t_k}; \boldsymbol{\lambda}^{*(k)})\left(\phi_{t_k}(X_{t_k})-\phi_{t_{k'}}(X_{t_{k'}})\right)\right\}=0.
\end{align}
As a consequence,
\begin{eqnarray*}
    &&E\left\{\widetilde{\phi}_{t_k}(X_{t_k}; \boldsymbol{\lambda}^{*(k)})\left(\widetilde{\phi}_{t_k}(X_{t_k}; \boldsymbol{\lambda}^{*(k)})-\phi_{t_{k'}}(X_{t_{k'}})\right)\right\}\\
    &=&E\left\{\widetilde{\phi}_{t_k}(X_{t_k}; \boldsymbol{\lambda}^{*(k)})\left( T_k^t\left(\phi_{t_{k}}(X_{t_{k}})-\phi_{t_{1}}(X_{t_{1}}), \dots, \phi_{t_{k}}(X_{t_{k}})-\phi_{t_{k-1}}(X_{t_{k-1}}), \phi_{t_{k}}(X_{t_{k}})\right)^t-\phi_{t_{k'}}(X_{t_{k'}})\right)\right\}\\
    &=&E\left\{\widetilde{\phi}_{t_k}(X_{t_k}; \boldsymbol{\lambda}^{*(k)})\left( T_k^t\left(\phi_{t_{k}}(X_{t_{k}})-\phi_{t_{1}}(X_{t_{1}}), \dots, \phi_{t_{k}}(X_{t_{k}})-\phi_{t_{k-1}}(X_{t_{k-1}}), \phi_{t_{k}}(X_{t_{k}})-\phi_{t_{k'}}(X_{t_{k'}})\right)^t\right)\right\}\\
    &=&0,
\end{eqnarray*}
where the second equation follows from the fact that the last element of $T_k$ equals 1 and the last equation from Equation \eqref{eq:proof1}.
Then,
\begin{eqnarray*}
    &&E\left\{\widetilde{\phi}_{t_k}(X_{t_k}; \boldsymbol{\lambda}^{*(k)})\left(\widetilde{\phi}_{t_k}(X_{t_k}; \boldsymbol{\lambda}^{*(k)})-\widetilde{\phi}_{t_{k'}}(X_{t_{k'}}; \boldsymbol{\lambda}^{*(k')})\right)\right\}\\
    &=&E\left\{\widetilde{\phi}_{t_k}(X_{t_k}; \boldsymbol{\lambda}^{*(k)})\left( \widetilde{\phi}_{t_k}(X_{t_k}; \boldsymbol{\lambda}^{*(k)})\vphantom{\left(\phi_{t_{k'}}(X_{t_{k'}})-\phi_{t_{1}}(X_{t_{1}}), \dots, \phi_{t_{k'}}(X_{t_{k'}})-\phi_{t_{k'-1}}(X_{t_{k'-1}}), \phi_{t_{k'}}(X_{t_{k'}})\right)^t}\right.\right.\\
    &&\left.\left.-T_{k'}^t\left(\phi_{t_{k'}}(X_{t_{k'}})-\phi_{t_{1}}(X_{t_{1}}), \dots, \phi_{t_{k'}}(X_{t_{k'}})-\phi_{t_{k'-1}}(X_{t_{k'-1}}), \phi_{t_{k'}}(X_{t_{k'}})\right)^t\right)\right\}\\
    &=&E\left\{\widetilde{\phi}_{t_k}(X_{t_k}; \boldsymbol{\lambda}^{*(k)})\left( \widetilde{\phi}_{t_k}(X_{t_k}; \boldsymbol{\lambda}^{*(k)})-\phi_{t_{k'}}(X_{t_{k'}})\vphantom{\left(\phi_{t_{k'}}(X_{t_{k'}})-\phi_{t_{1}}(X_{t_{1}}), \dots, \phi_{t_{k'}}(X_{t_{k'}})-\phi_{t_{k'-1}}(X_{t_{k'-1}}), \phi_{t_{k'}}(X_{t_{k'}})\right)^t}\right.\right.\\
    &&\left.\left.-T_{k'}^t\left(\phi_{t_{k'}}(X_{t_{k'}})-\phi_{t_{1}}(X_{t_{1}}), \dots, \phi_{t_{k'}}(X_{t_{k'}})-\phi_{t_{k'-1}}(X_{t_{k'-1}}), 0\right)^t\right)\right\}\\
    &=&E\left\{\widetilde{\phi}_{t_k}(X_{t_k}; \boldsymbol{\lambda}^{*(k)})\left( -T_{k'}^t\left(\phi_{t_{k'}}(X_{t_{k'}})-\phi_{t_{1}}(X_{t_{1}}), \dots, \phi_{t_{k'}}(X_{t_{k'}})-\phi_{t_{k'-1}}(X_{t_{k'-1}}), 0\right)^t\right)\right\}\\
    &=&E\left\{\widetilde{\phi}_{t_k}(X_{t_k}; \boldsymbol{\lambda}^{*(k)})\left(T_{k'}^t\vphantom{\left(\phi_{t_{k'}}(X_{t_{k'}})-\phi_{t_{1}}(X_{t_{1}}), \dots, \phi_{t_{k'}}(X_{t_{k'}})-\phi_{t_{k'-1}}(X_{t_{k'-1}}), \phi_{t_{k'}}(X_{t_{k'}})\right)^t}\left(\phi_{t_{k}}(X_{t_{k}})-\phi_{t_{k'}}(X_{t_{k'}})-(\phi_{t_{k}}(X_{t_{k}})-\phi_{t_{1}}(X_{t_{1}})), \dots,\right.\right.\right.\\
    &&\left.\left.\left.\phi_{t_{k}}(X_{t_{k}})-\phi_{t_{k'}}(X_{t_{k'}})-(\phi_{t_{k}}(X_{t_{k}})-\phi_{t_{k'-1}}(X_{t_{k'-1}})), 0\right)^t\right)\right\}\\
    &=&0,
\end{eqnarray*}
where the second equation follows from the fact that the last element of $T_k$ equals 1, the third equation from $E\left\{\widetilde{\phi}_{t_k}(X_{t_k}; \boldsymbol{\lambda}^{*(k)})\left(\widetilde{\phi}_{t_k}(X_{t_k}; \boldsymbol{\lambda}^{*(k)})-\phi_{t_{k'}}(X_{t_{k'}})\right)\right\}=0$ and the last equation from Equation \eqref{eq:proof1}.




\subsection*{Efficiency Improvement (or No Change)}
In this part of the proof, our goal is to show that the asymptotic variance is not increased by using $\widetilde\theta_{t_k}$ instead of $\widehat\theta_{t_k}$. 
First, notice that the asymptotic covariance matrix of $(\widehat\theta_{t_k}-\widehat\theta_{t_1}, \dots, \widehat\theta_{t_k}-\widehat\theta_{t_{k-1}}, \widehat\theta_{t_k})^t$ is equal to $Var(\widehat{\theta}_{t_k})$ times the covariance matrix $\boldsymbol\Sigma^{(k)}$ of  $(Z^*_k-(f_k/f_{1})^{1/2}Z^*_{1},\dots,Z^*_k-(f_k/f_{k-1})^{1/2}Z^*_{k-1}, Z^*_k)^t$. It is therefore sufficient to prove that $Var(\widetilde\theta_{t_k})$ is equal to or smaller than $Var(\widehat{\theta}_{t_k})\boldsymbol\Sigma^{(k)}_{k,k}$, with $\boldsymbol\Sigma^{(k)}_{k,k}$ element $(k,k)$ of  $\boldsymbol\Sigma^{(k)}$.

To this end, we consider a different representation of the updated estimator $\widetilde\theta_{t_k}$. In particular, $\widetilde\theta_{t_k}$ can be written as a linear transformation of the original estimators $\widehat\theta_{t_1}, \dots, \widehat\theta_{t_k}$. 
We assume that $\boldsymbol\Sigma^{(k)}$ is positive definite\footnote{If not, then one of the matrix elements is a linear combination of the others, and can be removed. This process can be repeated until all elements are linearly independent.}.
Then, the transformation matrix is a function of the Cholesky decomposition matrix $A^{(k)}$ of $\boldsymbol\Sigma^{(k)}$; i.e., $A^{(k)}$ is upper triangular and satisfies $\boldsymbol\Sigma^{(k)} = (A^{(k)})^tA^{(k)}$. In particular, we have that $\boldsymbol{\lambda}^{*(k)}=\{(\widetilde{A}^{(k)})^t\widetilde{A}^{(k)}\}^{-1}(\widetilde{A}^{(k)})^tA^{(k)}_k$, where $\widetilde{A}^{(k)}$ denotes the first $k-1$ columns of $A^{(k)}$ and $A^{(k)}_k$ the $k$th column of $A^{(k)}$. 
Then,
\begin{align}\label{eq:linTrans}
\widetilde{\theta}_{t_k} = \left[ (A^{(k)})^{-1} \left\{I_k - \widetilde{A}^{(k)} \{(\widetilde{A}^{(k)})^t\widetilde{A}^{(k)}\}^{-1}(\widetilde{A}^{(k)})^t \right\} A^{(k)}_k \right]^t W^{(k)},
\end{align}
where $W^{(k)}=(\widehat\theta_{t_k}-\widehat\theta_{t_1}, \dots, \widehat\theta_{t_k}-\widehat\theta_{t_{k-1}}, \widehat\theta_{t_k})^t$ and $I_k$ denotes the identity matrix with $k$ rows. We denote the linear transformation matrix $\left[ (A^{(k)})^{-1} \left\{I_k - \widetilde{A}^{(k)} \{(\widetilde{A}^{(k)})^t\widetilde{A}^{(k)}\}^{-1}(\widetilde{A}^{(k)})^t \right\} A^{(k)}_k \right]$ as $T_k$. Consequently,
\begin{eqnarray*}
    Var(\widetilde\theta_{t_k})&=&Var((T_k)^tW^{(k)})\\
    &=&(T_k)^tVar(W^{(k)})T_k\\
    &=&(T_k)^tVar(\widehat{\theta}_{t_k})\boldsymbol\Sigma^{(k)}T_k\\
    &=&Var(\widehat{\theta}_{t_k})\left[ (A^{(k)})^{-1} \left\{I_k - \widetilde{A}^{(k)} \{(\widetilde{A}^{(k)})^t\widetilde{A}^{(k)}\}^{-1}(\widetilde{A}^{(k)})^t \right\} A^{(k)}_k \right]^t(A^{(k)})^tA^{(k)}\\
    &&\left[ (A^{(k)})^{-1} \left\{I_k - \widetilde{A}^{(k)} \{(\widetilde{A}^{(k)})^t\widetilde{A}^{(k)}\}^{-1}(\widetilde{A}^{(k)})^t \right\} A^{(k)}_k \right]\\
    &=&Var(\widehat{\theta}_{t_k})\left[ \left\{I_k - \widetilde{A}^{(k)} \{(\widetilde{A}^{(k)})^t\widetilde{A}^{(k)}\}^{-1}(\widetilde{A}^{(k)})^t \right\} A^{(k)}_k \right]^t\\
    &&\left[ \left\{I_k - \widetilde{A}^{(k)} \{(\widetilde{A}^{(k)})^t\widetilde{A}^{(k)}\}^{-1}(\widetilde{A}^{(k)})^t \right\} A^{(k)}_k \right]\\
    &=&Var(\widehat{\theta}_{t_k})(A^{(k)}_{k,k})^2,
\end{eqnarray*}
with $A^{(k)}_{k,k}$ element $(k,k)$ of $A^{(k)}$. 
By the definition of the Cholesky decomposition $(A^{(k)}_{k,k})^2$ equals $\boldsymbol\Sigma^{(k)}_{k,k}-\sum_{j=1}^{k-1}(A^{(k)}_{k,j})^2$, which is equal to or smaller than $\boldsymbol\Sigma^{(k)}_{k,k}$. As a consequence, the estimator $\widetilde{\theta}_{t_k}$ at each analysis $k$ is asymptotically as or more precise as the original estimator $\widehat{\theta}_{t_k}$.

\subsection*{Monotonicity of $\widetilde{\mathcal{I}}^*_{k}$ and $\widetilde{\mathcal{I}}_{k}$}
The orthogonalization moreover ensures that $\widetilde{\mathcal{I}}^*_{k}$ is non-decreasing over analysis times $t_1, \dots, t_K$. We can prove this by contradiction. Specifically, for $t_{k'}<t_k$ if $\widetilde{\mathcal{I}}^*_{k}<\widetilde{\mathcal{I}}^*_{k'}$ then the (asymptotic) variance in Equation \eqref{eq:minimizer} is not minimized as choosing $(\lambda^{*(k)}_{1}, \dots, \lambda^{*(k)}_{k'-1})=(\lambda^{*(k')}_{1}, \dots, \lambda^{*(k')}_{k'-1})$, $\lambda^{*(k)}_{k'}=1-\sum_{j=1}^{k'-1}\lambda^{*(k')}_{j}$ and $(\lambda^{*(k)}_{k'+1}, \dots, \lambda^{*(k)}_{k-1})=(0, \dots, 0)$ would lead to a smaller variance (i.e., the variance of $\widetilde\theta_{t_{k'}}$). A similar reasoning holds for finite samples by replacing $\boldsymbol{\lambda}^{*(k)}$ and $\boldsymbol{\lambda}^{*(k')}$ by respectively $\widehat{\boldsymbol{\lambda}}^{(k)}$ and $\widehat{\boldsymbol{\lambda}}^{(k')}$.

\section{Additional Information on Estimators Used for\\ Simulations}
In this Appendix we describe the different estimators used for the estimation of $\theta=E\left(Y|A=1\right)-E\left(Y|A=0\right)$ in the simulation studies in Section~\ref{sec:dataAnal} of the main article.

\subsection{Unadjusted Estimator}
For a given population, the unadjusted estimator of the average treatment effect is the difference between sample means of $Y$ comparing those assigned to $A = 1$ versus $A = 0$:
$$\widehat\theta_{unadj}=\frac{\sum_{i=1}^nA_iY_i}{\sum_{i=1}^nA_i}-\frac{\sum_{i=1}^n(1-A_i)Y_i}{\sum_{i=1}^n(1-A_i)}.$$
This estimator is the unadjusted estimator used at the final analysis.

For the interim analyses, we assume that missing data is only caused by administrative censoring due to some participants not having experienced the final outcome at time $t$.
The restriction of the full data to the data collected up to any calendar time $t$ is represented by $X_{t,i}$. Let $C^{(0)}_{i,t}$ and $C^{(1)}_{i,t}$ denote the indicator whether respectively the baseline measurements $(E, W, A)$ and outcome $Y$ are observed at time $t$ for participant $i$. Note that $\Delta_i(t)=C^{(0)}_{i,t}$. At time $t$, we can then distinguish three cohorts of patients: a first cohort of patients for whom all data are available $((C^{(0)}_{i,t}, C^{(1)}_{i,t}) = (1, 1))$, a second cohort of patients who are enrolled (i.e., $(E, W, A)$ is observed) but for whom the outcome $Y$ is not yet observed $((C^{(0)}_{i,t}, C^{(1)}_{i,t}) = (1, 0))$, and a third cohort of patients who are not yet enrolled and thus for whom no data are observed $((C^{(0)}_{i,t}, C^{(1)}_{i,t}) = (0, 0))$. 

The pipeline participants at time $t$ are those patients enrolled but with $Y$ not yet observed (i.e., $((C^{(0)}_{i,t}, C^{(1)}_{i,t}) = (1, 0))$). For a given population, the unadjusted estimator of the average treatment effect at time $t$ is the difference between sample means of $Y$ comparing those assigned to $A = 1$ versus $A = 0$ among the participants who have $Y$ observed at time $t$:
$$\widehat\theta_{unadj,t}=\frac{\sum_{i=1}^nC^{(1)}_{i,t}A_iY_i}{\sum_{i=1}^nC^{(1)}_{i,t}A_i}-\frac{\sum_{i=1}^nC^{(1)}_{i,t}(1-A_i)Y_i}{\sum_{i=1}^nC^{(1)}_{i,t}(1-A_i)}.$$
This estimator is the unadjusted estimator used at the interim analyses and during monitoring of the data.

\subsection{Standardization Estimator}\label{sec:covAdj}
To take full advantage of prognostic baseline covariates, we focus on a generalization of the covariate adjusted estimator in \cite{ge2011covariate}, which was suggested in the recent FDA guidance on covariate adjustment \cite{FDA2021}. Importantly, this covariate adjusted estimator is robust to model misspecification and has a high potential to improve precision \citep{benkeser2020improving}, making it a low-risk, high-reward method. 

For a binary endpoint, a covariate adjusted estimator for the estimand of interest, $\theta$, can be obtained at time $t$ as follows:
\begin{enumerate}
    \item Fit a logistic regression model with maximum likelihood that regresses the outcome $Y$ on prespecified baseline covariates $W$ among the patients with $A=1$ and $C^{(1)}_{t}=1$. The model should include an intercept term.
    \item For each participant with $C^{(0)}_{t}=1$, use the fitted regression model in Step 1 to compute a prediction of the probability of response under $A=1$; and take the average of these predicted probabilities to obtain an estimator for the average response under $A=1$.
    \item Fit a logistic regression model with maximum likelihood that regresses the outcome $Y$ on prespecified baseline covariates $W$ among the patients with $A=0$ and $C^{(1)}_{t}=1$. The model should include an intercept term.
    \item For each participant with $C^{(0)}_{t}=1$, use the fitted regression model in Step 3 to compute a prediction of the probability of response under $A=0$; and take the average of these predicted probabilities to obtain an estimator for the average response under $A=0$.
    \item Take the difference of the estimates of the average response rate under $A=1$ (Step 2) and $A=0$ (Step 4), in order to obtain an estimate $\widehat\theta_{stand,t}$ of the marginal risk difference $\theta$.
\end{enumerate}
This estimator is the standardization estimator used at the interim analyses and to determine the timing of the interim analyses (for the standardization estimator) when monitoring the data.

At the final analysis, the standardization estimator $\widehat\theta_{stand}$ will be based on predictions (see Step 2 and 4 in the algorithm above) for all participants as $C^{(1)}_{i, t}=1$,  $\forall i=1, \dots, n$. Correspondingly, at any time $t$, we denote by $\widehat\theta^*_{stand, t}$ the standardization estimator at time $t$ where we only make predictions for patients with $C^{(1)}_{t}=1$ (and not for the pipeline patients with $C^{(1)}_{t}=0$ and $C^{(0)}_{t}=1$).
This estimator is the standardization estimator used at the final analysis and to determine the timing of the final analysis (for the covariate adjusted estimator) when monitoring the data (Aim (a)) or updating $n_{max}$ at the interim analyses (Aim (b)).

The variance (and covariance) of the estimators described above can be obtained by the nonparametric bootstrap or a sandwich estimator via the Delta method. 

\subsection{Targeted Maximum Likelihood Estimator}\label{sec:tmle}

The Longitudinal Targeted Maximum Likelihood Estimator (TMLE) is an estimator of the average treatment effect that makes use of intermediate outcomes in longitudinal studies assessed at study visits between randomization and the visit where primary outcome is obtained. Intermediate outcomes are used in modeling a censoring mechanism and in constructing a sequence of outcome regression fits that are inversely weighted according to propensity models for treatment and censoring.

Let $j = 1, \ldots, J+1$ indicate the study visits after randomization, and $R_{ij}$ indicate whether $Y_{ij}$, the outcome for participant $i$ at study visit $j$ is observed. At each visit, let $H_{ij}$ denote the event history for participant $i$ prior to visit $j$, i.e. all baseline covariates, treatment assignment, and all prior outcomes: $H_{j} = (W, A, Y_{1}, \ldots, Y_{(j-1)})$.

Outcome data are assumed to follow a monotone missingness pattern, i.e. that $R_{ij} = 0$ implies $R_{ij'} = 0$ for $j' > j$. Any missing outcomes that do not follow a monotone missingness pattern are referred to as intermittent missing values, and must be imputed.

The longitudinal TMLE is constructed in two steps. First, a sequence of weight variables is constructed:

\begin{enumerate}
  \item Impute intermittent missing values: At each visit $j$, construct an appropriate regression model for imputing intermittent values based on the event history $H_{j}$. After imputation, data should follow a monotone missingness pattern.
  \item Fit a propensity score model for treatment assignment: this is a logistic regression model with an intercept term, estimated using maximum likelihood, that regresses the treatment indicator $A$ on the pre-specified baseline covariates $W$: $\hat{\psi} = Pr\{A = 1 \vert W\}$. 
  \item Use the propensity score model for treatment assignment to compute each individual's predicted probability of receiving their observed treatment level, indicated by $a\in \{0, 1\}$, based on their pre-randomization covariates: Let $\pi_{i}^{A} = (\hat{\psi_{i}})^{a}(1-\hat{\psi}_{i})^{(1-a)}$.
  \item Iterating from the first to the final post-randomization visit, fit a propensity model for being uncensored at that visit based on their event history prior to that visit given that they were not already censored prior to the study visit. This is a logistic regression model with an intercept term, estimated using maximum likelihood, that regresses the observed data indicator $R_{j}$ on $H_{j}$ among those with $R_{(j-1)} = 1$: Let $\pi_{ij}^{R} = Pr\{R_{j} = 1 \vert H_{j}, R_{(j-1)} = 1\}$ denote the fitted probability of being observed at visit $j$.
  \item Construct a weight variable $N_{j}$ for each outcome. This variable is missing for those whose outcomes were not observed at the previous study visit, and otherwise it is set to the inverse of the probability of receiving the observed treatment multiplied by the inverse of the cumulative probability of being uncensored up to visit $j$:
  
  $$N_{ij} = \frac{1}{\pi_{i}^{A}}\prod_{j' = 1}^{j} \frac{1}{\pi_{ij'}^{R}}$$
\end{enumerate}

Next, a sequence of variables $Q_{(J+1)}, \ldots, Q_{1}$ are constructed using regression models that are weighted by the variables constructed in the first steps. This process iterates back from the final outcome to the first study visit. For the final study visit, define $Q_{(J+1)} = Y_{(J+1)}$. Iterating backwards from $j = J, \ldots, 0$, with $j=0$ indicating the visit at baseline:

\begin{enumerate}
  \item Regress $Q_{(j+1)}$ on $H_{(j+1)}$ using a weighted linear (for continuous) or logistic (for binary) regression, using $N_{(j + 1)}$ as the weighting variable in the subset of participants with $R_{(j + 1)} = 1$. Denote this regression model as $\mathcal{M}_{(j + 1)}(H_{(j + 1)})$.
  
  \item For participants with $R_{j} = 1$ (i.e. those with no missing values in $H_{(j + 1)}$), set $Q_{j}$ to the fitted value from this weighted regression.
\end{enumerate}

The iterative process creates the model $\mathcal{M}_{1}(W, A)$. Let $Q_{i0}^{a'}$ denote the fitted value from $\mathcal{M}_{1}(W_{i}, A')$ for each individual $i$ when $A'$ is set to $a' \in \{0, 1\}$. The TMLE estimator is given by: 

$$\hat{\theta}_{TMLE} = \frac{1}{n}\sum_{i=1}^{n}\left({Q_{i0}^{1}} - {Q_{i0}^{0}}\right).$$

\section{Computing Efficacy Boundaries}\label{app:bound}
In this section, we describe how to compute the efficacy stopping boundary $c_k$ at analysis $k$. The goal is to do this in a way that guarantees (asymptotically) a specified Type I error $\alpha$.

We estimate the information fraction, denoted as $\pi_{t_k}$, by  $\widetilde{\mathcal{I}}_{k}=(\widehat{se}(\widetilde\theta_{t_k}))^{-2}$ divided by the maximum information (defined in Section~\ref{subsec:inf_impl}), and find the boundary $c_k$ that (asymptotically) solves the equation
$$P\left(|\widetilde Z_1|\leq c_1, \dots, |\widetilde Z_{k-1}|\leq c_{k-1}, |\widetilde Z_k|\geq c_k\right)=\alpha(\pi_k)-\alpha(\pi_{k-1}),$$
with $\alpha(\pi)$ an  error spending function, where $0\leq \pi\leq 1$. The commonly used \cite{o1979multiple} boundaries, for example,  can be approximated by using an appropriate choice of the error spending function  \citep{gordon1983discrete}. This can be computed  with the \texttt{ldBounds} function in the \texttt{R} package \texttt{ldbounds} by passing the information fraction for the times $t_1, \dots, t_k$, the total Type I error $\alpha$ and the type of error spending function.

If $\widetilde Z_k> c_k$ ($k=1, \dots, K$) then the trial is stopped and the null hypothesis is rejected, otherwise the trial is continued to the next monitoring time. If the test continues to the $K$th analysis, the null hypothesis is rejected if $\widetilde Z_K>c_K$. Otherwise, if $\widetilde Z_k \leq c_k$ at all times $k$ then we fail to reject the null hypothesis.

\section{Additional Simulation Results Based on MISTIE III Trial}\label{app:addSim}
\subsection{Results for Larger Sample Size}\label{app:addSim_large}
The results in this section (see Table \ref{tab:data_InfMonLarge} and \ref{tab:data_sim_gsdLarge}) are obtained under the same simulation design as in Section~\ref{sec:sim} of the main article but with a target treatment effect $\theta_A$ of 0.065 (instead of 0.13). For a trial without interim analyses, this results in a maximum information of 2327 and a maximum sample size of 1882. For the latter one, we assumed that the probability of a successful outcome in the control arm equals 0.25. For trials with one interim analysis at information fraction 0.50, we need a maximum information of 2591 and a maximum total sample size of 2096. In the latter case, the interim analysis is conducted when 1048 patients have their primary endpoint observed.
The uniform recruitment rate of approximately 12 patients per month corresponds with the average recruitment rate in the MISTIE III Trial.
For the data generating mechanisms described above, we perform $25,000$ Monte Carlo runs under the null hypothesis.
\spacingset{1} 
\begin{table}[h]
	\caption{\label{tab:data_InfMonLarge} Results for information adaptive design and maximum sample size trials (without interim analysis) for $\theta_A=0.065$. 
	}
\centering
\begin{tabular}{ l  l c  c c c }
  \hline
 Design Type&& Type I & \textbf{ASN} & AAT & AI\\
  \hline  
    && \multicolumn{4}{ c }{$\theta=0$ (Null)}\\
  \cline{3-6} 
  Information Adaptive & Unadj. &5.01\%  & \textbf{2280} & 6401& 2328\\
  with $\mathcal{I}(\theta_A)=2327$ & Stand. &5.12\% & \textbf{1528} & 4407 & 2171\\
   & TMLE &4.79\% & \textbf{1506} & 4350 & 2172\\ \\
Maximum Sample Size& Unadj. &4.94\% & - & 5347 & 1921\\
   with $n_{max}=1882$ & Stand. &5.12\% & - & 5347 & 2676 \\
   & TMLE &5.07\% & - & 5346 & 2716\\ \\
   \hline
\end{tabular}\\
	 {\raggedright Unadj., unadjusted estimator; Stand., standardization estimator; TMLE, targeted maximum likelihood estimator; ASN, average sample number; AAT, average analysis time (in days); AI, average information. \par}
\end{table}
\spacingset{1.85} 
The simulation results in Table \ref{tab:data_InfMonLarge} show that the Type 1 error of both information adaptive and maximum sample size trials are (approximately) preserved, and that it is closer to the nominal level of 5\% than for the results under smaller sample sizes and maximum information levels (see Table 1 in the main article). Similar results can be seen for group sequential designs with one interim analysis in Table \ref{tab:data_sim_gsdLarge}.
\spacingset{1} 
\begin{table}[h!]
\caption{\label{tab:data_sim_gsdLarge} Results for group sequential designs with and without adaptive analysis timing for $\theta_A=0.065$.
	}
\centering
\begin{tabular}{ l  l   c c c c  c cc}
  \hline
  Design Type&& Type I & \textbf{ASN} & ASN1 & ASN2  & AAT& AAT1 & AAT2  \\
  \hline\\
      && \multicolumn{7}{ c }{$\theta=0$ (Null)}\\
  \cline{3-9} 
    Max. Sample Size & Unadj. &5.04\% & \textbf{2069} & 1186 & 2096   &5832 & 3139 & 5914 \\
    with $\widehat\theta_{t_k}$& Stand. &5.32\% &  \textbf{2064}& 1186 & 2096   & 5815&  3139&  5914\\
    & TMLE &4.96\%   & \textbf{2065}& 1186 & 2096 & 5818 & 3140 & 5913\\
    Max. Sample Size & Stand. & 5.35\% & \textbf{2064} & 1186 & 2096  & 5815& 3139 & 5914 \\
    with $\widetilde\theta_{t_k}$& TMLE &4.98\%   & \textbf{2065}& 1186 & 2096 & 5818 & 3140 & 5913\\ \\
    Information Adaptive & Unadj. & 4.99\% & \textbf{2504} & 1412 & 2539 & 6983 & 3736 & 7087  \\
    with $\widehat\theta_{t_k}$& Stand. & 5.27\% & \textbf{1671} & 1002 & 1694  & 4774& 2652 & 4846 \\
    & TMLE & 5.05\%   & \textbf{1649}& 964 & 1671 & 4718 & 2552 &4785 \\
    Information Adaptive & Stand. & 5.25\%  & \textbf{1671}  & 1002 & 1694 &4774 & 2652  & 4846  \\
    with $\widetilde\theta_{t_k}$& TMLE & 5.01\%   & \textbf{1649}& 964 & 1671 & 4718 & 2552 &4785 \\\\ 
\hline
\end{tabular}
	{\raggedright Unadj., unadjusted estimator; Stand., standardization estimator; TMLE, targeted maximum likelihood estimator; ASN, average sample number; ASNj, average sample number at analysis $j$ ($j=1,2$); AAT, average analysis time (in days); AATj, average analysis time (in days) of analysis $j$ ($j=1,2$). \par}
\end{table}
\spacingset{1.85} 
\newpage
\subsection{Results Using Small Sample Correction}\label{app:addSim_correction}
The results in this section (see Table \ref{tab:data_sim_InfMonCorrection} and \ref{tab:data_sim_gsdCorrection}) are obtained under the same simulation design as in Section~\ref{sec:sim} of the main article. The only difference is that at each analysis time, we use the variance estimator proposed in \cite{tsiatis2008covariate}, following ideas from \cite{koch1998issues} and \cite{lesaffre2003note}, to correct for the estimation of nuisance parameters in small samples. Specifically, we multiply the variance estimator by a small-sample `correction factor'. For the standardization estimator described in Section \ref{sec:covAdj}, we used the correction factor 
$$\frac{(n_0-p_0-1)^{-1}+(n_1-p_1-1)^{-1}}{(n_0-1)^{-1}+(n_1-1)^{-1}},$$
where $n_j$ ($j=0,1$) are the number of participants used to fit the outcome working model in treatment arm $j$ and $p_j$ the numbers of parameters fitted in these models, exclusive of intercepts. 
Furthermore, for the targeted maximum likelihood estimator described in Section \ref{sec:tmle}, we used the correction factor $(n-1)/(n-p-1)$,
where $n$ is the number of participants used to fit the outcome working model and $p$ the numbers of parameters fitted in that model, exclusive of intercepts.

The results in Tables \ref{tab:data_sim_InfMonCorrection} and \ref{tab:data_sim_gsdCorrection} show better performance with respect to the Type I error inflation compared to using standard Wald statistics without small sample correction. This is accompanied by a small loss of power for maximum sample size designs. The average sample number, average analysis time and average information are in line with previous results.
\spacingset{1} 
\begin{table}[h!]
	\caption{\label{tab:data_sim_InfMonCorrection} Results for information adaptive design and maximum sample size trials (without interim analysis) using small sample correction. 
	}
\centering
\begin{tabular}{ l  l c  c c c ccc cc}
  \hline
  && \multicolumn{9}{ c }{Simulation Parameter}\\
  && \multicolumn{4}{ c }{$\theta=0.13$ (Alternative)} && \multicolumn{4}{ c }{$\theta=0$ (Null)}\\
  \cline{3-6}  \cline{8-11} 
 Design Type&& Power & \textbf{ASN} & AAT & AI && Type I & \textbf{ASN} & AAT & AI\\
  \hline  
  Info. Adaptive & Stand. &87.4\%&\textbf{438} &1523 &563 & & 5.06\% &\textbf{406} &1438&563\\
  with $\mathcal{I}(\theta_A)=582$ & TMLE &86.9\%& \textbf{432}& 1506& 553& & 4.74\% &\textbf{402} &1428&556\\ \\
Max. Sample Size& Stand.&90.8\%&- &1682 &642 & & 4.94\% &- &1682 &694\\
   with $n_{max}=498$ & TMLE &91.3\%& -& 1682&642 & & 4.63\% & -&1682&694\\ \\
   Max. Sample Size & Stand. &94.3\%&- &1893 &749 & & 4.97\% &- &1894 &809\\
   with $n_{max}=578$ & TMLE &94.4\%&- &1893 &750 & & 4.78\% & -&1894&811\\ 
   \hline
\end{tabular}
 {\raggedright Stand., standardization estimator; TMLE, targeted maximum likelihood estimator; ASN, average sample number; AAT, average analysis time (in days); AI, average information. \par}
\end{table}

\begin{table}[H]
	\caption{\label{tab:data_sim_gsdCorrection} Results for group sequential designs with and without information adaptive analysis timing using small sample correction. 
	}
\centering
\begin{tabular}{ l  l   c c c c  c cc}
  \hline
  Design Type&& Power  & \textbf{ASN} & ASN1 & ASN2 & AAT& AAT1 & AAT2   \\
  \hline\\
  && \multicolumn{7}{ c }{$\theta=0.13$ (Alternative)}\\
  \cline{3-9} 
  Max. Sample Size & Stand. &90.9\% & \textbf{468} &  415& 554  & 1375&  1098&  1830\\
    with $\widehat\theta_{t_k}$& TMLE &91.0\%   &\textbf{456} & 415 & 554 & 1314 & 1098 &1830\\
    Max. Sample Size & Stand. & 90.9\% &  \textbf{468}& 415 & 554  & 1375&  1098&  1830\\
    with $\widetilde\theta_{t_k}$& TMLE &90.9\%   &\textbf{456} & 415 & 554 & 1314 & 1098 &1830\\ \\
   Information Adaptive & Stand. &87.1\% & \textbf{431} & 387 & 487  & 1299& 1024 & 1652 \\
    with $\widehat\theta_{t_k}$& TMLE &87.4\%  &\textbf{408} & 347 & 484 & 1240 & 917 &1644 \\ 
    Information Adaptive  & Stand. &87.0\% & \textbf{431} & 387 & 487  & 1299& 1024 & 1652 \\
    with $\widetilde\theta_{t_k}$& TMLE &87.4\%  &\textbf{408} & 347 & 484 & 1240 & 917 &1644 \\ \\
      && \multicolumn{7}{ c }{$\theta=0$ (Null)}\\
  \cline{3-9} 
    Max. Sample Size & Stand. &5.17\% & \textbf{549} & 415  & 554  & 1806& 1098 & 1831  \\
    with $\widehat\theta_{t_k}$& TMLE &5.09\%   & \textbf{549}& 415 & 554 & 1803 & 1098 & 1830\\
    Max. Sample Size & Stand. & 5.16\% & \textbf{549} & 415  & 554  & 1806&  1098&  1831\\
    with $\widetilde\theta_{t_k}$& TMLE &5.08\%   & \textbf{549}& 415 & 554 & 1803 & 1098 & 1830\\ \\
    Information Adaptive & Stand. & 5.06\%  & \textbf{449} & 370 & 452  & 1542& 978 & 1560\\
    with $\widehat\theta_{t_k}$& TMLE &4.42\%   &\textbf{446} & 328 & 449 & 1534 &867  &1551 \\ 
    Information Adaptive & Stand. & 5.05\%  & \textbf{449} & 370 & 452  & 1542& 978 & 1560 \\
    with $\widetilde\theta_{t_k}$& TMLE &4.40\%   &\textbf{446} & 328 & 449 & 1534 &867  &1551 \\\\ 
\hline
\end{tabular}
{\raggedright Stand., standardization estimator; TMLE, targeted maximum likelihood estimator; ASN, average sample number; ASNj, average sample number at analysis $j$ ($j=1,2$); AAT, average analysis time (in days); AATj, average analysis time (in days) of analysis $j$ ($j=1,2$). \par}
\end{table}
\spacingset{1.85} 

\newpage
\subsection{Results Under Different Prognostic Values}\label{app:addSim_prognostic}

The results in this section are obtained under the same simulation design as in Section~\ref{sec:sim} of the main article but with a different prognostic value of the baseline covariates.
In Table \ref{tab:data_InfMon_notPrognostic} and Table \ref{tab:data_sim_gsd_notprognostic}, we present the results under the setting where the baseline covariates are not prognostic (i.e., independent) for the primary outcome. This is obtained by first following the same resampling scheme as for the data-generating mechanisms explained in Section~\ref{sec:sim} of the main article. For these generated data vectors $(A, W, X_{30}, X_{180}, Y)$, we randomly replaced the baseline covariates $W$ by resampling from these initial data vectors. Similarly, the results in Table \ref{tab:data_InfMon_mediumPrognostic} and Table \ref{tab:data_sim_gsd_mediumprognostic} are obtained under the setting where the baseline covariates have a `medium' (i.e., less prognostic but not independent) prognostic value. In this case, we randomly replace $W$ with probability 0.50 as for the case where baseline covariates are independent of the primary outcome.

The simulation results under the null hypothesis ($\theta=0$) show that the Type 1 error is (approximately) preserved under the different designs.
The results moreover show that the information adaptive design trials -with and without interim analysis- achieve the desired power of 88\% under the alternative $\theta_A=0.13$ for the unadjusted estimator. However, for the covariate adjusted estimators, we see a small power loss of around 2\% and 1.5\% for respectively the non-prognostic and `medium' prognostic setting. This small loss in power does not outweigh the power gain or the decrease in the total number of participants when the covariates are prognostic. 
The reason for this power loss is two-fold. First, there is some power loss due to adjusting for noise, especially in the case where the covariates are not prognostic. Second, early on in the trial there is an underestimation of the variance (and overestimation of the information), this leads to a prediction of the analysis times and total sample numbers that's too small. By not allowing to restart recruitment once it has been stopped, this eventually results in too early analyses with too few participants. It's a topic for future research to investigate how to select prognostic variables (and not select non-prognostic variables) and how to optimize the information adaptive timing of the analyses. 
Similarly, for designs where the timing of the analysis is based on the sample number (i.e., maximum sample size designs with one (final) analysis in Tables \ref{tab:data_InfMon_notPrognostic} and \ref{tab:data_InfMon_mediumPrognostic} and maximum sample size GSDs in Tables \ref{tab:data_sim_gsd_notprognostic} and \ref{tab:data_sim_gsd_mediumprognostic}), we also observe a small power loss for the covariate adjusted estimators compared to the unadjusted estimators due to the loss of degrees of freedom.

Finally, when the baseline covariates are prognostic (see Tables \ref{tab:data_InfMon_mediumPrognostic} and \ref{tab:data_sim_gsd_mediumprognostic}), covariate adjustment leads to a gain in power when the sample numbers at the different analyses are fixed (see maximum sample size design with one (final) analysis in Table \ref{tab:data_InfMon_mediumPrognostic}  and maximum sample size GSD in Table \ref{tab:data_sim_gsd_mediumprognostic}) and to a reduced average analysis time and average sample number when the timing is information adaptive.
\spacingset{1} 
\begin{table}[H]
	\caption{\label{tab:data_InfMon_notPrognostic} Results for information adaptive design and maximum sample size trials (without interim analysis) for non-prognostic covariates. 
	}
\centering
\begin{tabular}{ l  l c  c c c ccc cc}
  \hline
  && \multicolumn{9}{ c }{Simulation Parameter}\\
  && \multicolumn{4}{ c }{$\theta=0.13$ (Alternative)} && \multicolumn{4}{ c }{$\theta=0$ (Null)}\\
  \cline{3-6}  \cline{8-11} 
 Design Type&& Power & \textbf{ASN} & AAT & AI && Type I & \textbf{ASN} & AAT & AI \\
  \hline  
  Info. Adaptive & Unadj. &88.7\%& \textbf{571} & 1875 &582  && 5.18\% & \textbf{569} & 1871 & 582 \\
  with $\mathcal{I}(\theta_A)=582$ & Stand. &86.1\%& \textbf{557} & 1838&563  && 5.03\%& \textbf{555} & 1833 & 564 \\
   & TMLE &86.8\%& \textbf{530} & 1766&567  && 5.13\% & \textbf{559} & 1843 & 565 \\ \\
Max. Sample Size& Unadj. &83.2\%& - & 1682&508  && 5.18\% & - & 1682 & 509 \\
   with $n_{max}=498$ & Stand.&82.3\%& - & 1682&502  && 5.01\%& - & 1682 & 504 \\
   & TMLE &84.6\%& - & 1682 &531 && 5.08\% & - & 1682  & 501 \\ 
   \hline
\end{tabular}
	 {\raggedright Unadj., unadjusted estimator; Stand., standardization estimator; TMLE, targeted maximum likelihood estimator; ASN, average sample number; AAT, average analysis time (in days); AI, average information. \par}
\end{table}

\begin{table}[H]
	\caption{\label{tab:data_sim_gsd_notprognostic} Results for group sequential designs with and without information adaptive analysis timing for non-prognostic covariates. 
	}
\centering
\begin{tabular}{ l  l   c c c c  c cc}
  \hline
  Design Type&& Power & \textbf{ASN} & ASN1 & ASN2  & AAT& AAT1 & AAT2  \\
  \hline\\
  && \multicolumn{7}{ c }{$\theta=0.13$ (Alternative)}\\
  \cline{3-9} 
  Max. Sample Size & Unadj. & 82.7\% & \textbf{484} & 415 & 554  & 1463& 1098 & 1830   \\
    with $\widehat\theta_{t_k}$& Stand. & 82.2\% & \textbf{487} & 415 & 554  & 1477& 1098 & 1830   \\
    & TMLE & 82.4\%  & \textbf{472} & 415 & 554 & 1395& 1098 & 1830   \\
    Max. Sample Size & Stand. & 82.4\%  & \textbf{487} & 415 & 554 & 1477& 1098 & 1830   \\
    with $\widetilde\theta_{t_k}$& TMLE & 82.2\%  & \textbf{472} & 415 & 554 & 1395& 1098 & 1830   \\ \\
   Information Adaptive & Unadj. & 88.4\%  & \textbf{534} & 461 & 636 & 1564& 1218 & 2046   \\
    with $\widehat\theta_{t_k}$& Stand. & 86.5\%  & \textbf{525} & 456 & 617 & 1545& 1206 & 1995   \\
    & TMLE & 87.5\% & \textbf{499} & 399 & 632  & 1474& 1054 & 2035   \\
    Information Adaptive  & Stand. & 86.4\%  & \textbf{525} & 456 & 617  & 1545& 1206 & 1995  \\
    with $\widetilde\theta_{t_k}$& TMLE & 87.4\% & \textbf{499} & 399 & 632  & 1474& 1054 & 2035  \\ \\
      && \multicolumn{7}{ c }{$\theta=0$ (Null)}\\
  \cline{3-9} 
    Max. Sample Size & Unadj. & 4.99\% & \textbf{550} & 415 & 554  & 1809& 1098 & 1830   \\
    with $\widehat\theta_{t_k}$& Stand. & 4.99\% & \textbf{550} & 415 & 554  & 1809& 1098 & 1830   \\
    & TMLE & 5.00\%  & \textbf{549} & 415 & 554  & 1806& 1098 & 1830  \\
    Max. Sample Size & Stand. & 4.99\%  & \textbf{550} & 415 & 554  & 1809& 1098 & 1830  \\
    with $\widetilde\theta_{t_k}$& TMLE &4.97\%  & \textbf{549} & 415 & 554  & 1806& 1098 & 1830  \\\\
    Information Adaptive & Unadj. & 5.26\% & \textbf{628} & 459 & 634   & 2014& 1215 & 2042  \\
    with $\widehat\theta_{t_k}$& Stand. & 5.07\%  & \textbf{609} & 455 & 614 & 1963& 1204 & 1989   \\
    & TMLE & 5.28\%  & \textbf{621} & 390 & 629 & 1995& 1031 & 2029   \\
    Information Adaptive & Stand. & 5.08\% & \textbf{609} & 455 & 614  & 1963& 1204 & 1989   \\
    with $\widetilde\theta_{t_k}$& TMLE & 5.29\%  & \textbf{621} & 390 & 629 & 1995& 1031 & 2029  \\\\ 
\hline
\end{tabular}
	{\raggedright Unadj., unadjusted estimator; Stand., standardization estimator; TMLE, targeted maximum likelihood estimator; ASN, average sample number; ASNj, average sample number at analysis $j$ ($j=1,2$); AAT, average analysis time (in days); AATj, average analysis time (in days) of analysis $j$ ($j=1,2$). \par}
\end{table}

\begin{table}[H]
	\caption{\label{tab:data_InfMon_mediumPrognostic} Results for information adaptive design and maximum sample size trials (without interim analysis) under `medium' prognostic value of the covariates. 
	}
\centering
\begin{tabular}{ l  l c  c c c ccc cc}
  \hline
  && \multicolumn{9}{ c }{Simulation Parameter}\\
  && \multicolumn{4}{ c }{$\theta=0.13$ (Alternative)} && \multicolumn{4}{ c }{$\theta=0$ (Null)}\\
  \cline{3-6}  \cline{8-11} 
 Design Type&& Power & \textbf{ASN} & AAT & AI  && Type I & \textbf{ASN} & AAT & AI\\
  \hline  
  Info. Adaptive & Unadj. &88.3\%& \textbf{571} & 1875&582  && 5.22\% & \textbf{569} & 1871 & 582 \\
  with $\mathcal{I}(\theta_A)=582$ & Stand. &86.6\%& \textbf{528} & 1760&566  && 5.14\%& \textbf{518} & 1735 & 565 \\
   & TMLE &86.8\%& \textbf{530} & 1766&567  && 5.12\%&  \textbf{521} & 1741 & 567 \\ \\
Max. Sample Size& Unadj. &83.0\%& - & 1682&508  && 5.14\%& - & 1682 & 509 \\
   with $n_{max}=498$ & Stand.&84.7\%& - & 1682&533  && 5.11\%& - & 1682 & 542 \\
   & TMLE &84.6\%& - & 1682&531  && 4.78\%& - & 1682 & 541 \\ 
   \hline
\end{tabular}
	 {\raggedright Unadj., unadjusted estimator; Stand., standardization estimator; TMLE, targeted maximum likelihood estimator; ASN, average sample number; AAT, average analysis time (in days); AI, average information. \par}
\end{table}

\begin{table}[H]
	\caption{\label{tab:data_sim_gsd_mediumprognostic} Results for group sequential designs with and without information adaptive analysis timing under `medium' prognostic value of the covariates. 
	}
\centering
\begin{tabular}{ l  l   c c c c  c cc}
  \hline
  Design Type&& Power & \textbf{ASN} & ASN1 & ASN2  & AAT& AAT1 & AAT2   \\
  \hline\\
  && \multicolumn{7}{ c }{$\theta=0.13$ (Alternative)}\\
  \cline{3-9} 
  Max. Sample Size & Unadj. & 83.0\% & \textbf{485} & 415 & 554  & 1468& 1098 & 1830   \\
    with $\widehat\theta_{t_k}$& Stand. & 84.5\% & \textbf{483} & 415 & 554  & 1456& 1098 & 1830   \\
    & TMLE & 84.6\% & \textbf{468} & 415 & 554  & 1378& 1098 & 1830   \\
    Max. Sample Size & Stand. & 84.4\%  & \textbf{483} & 415 & 554 & 1456& 1098 & 1830   \\
    with $\widetilde\theta_{t_k}$& TMLE& 84.4\%  & \textbf{468} & 415 & 554 & 1378& 1098 & 1830   \\ \\
   Information Adaptive & Unadj. & 88.7\%  & \textbf{535} & 460 & 636 & 1567& 1217 & 2046   \\
    with $\widehat\theta_{t_k}$& Stand. & 86.6\% & \textbf{502} & 439 & 584  & 1486& 1161 & 1910   \\
    & TMLE & 87.2\%  & \textbf{477} & 386 & 595 & 1416& 1020 & 1938   \\
    Information Adaptive  & Stand. & 86.5\% & \textbf{502} & 439 & 584  & 1486& 1161 & 1910   \\
    with $\widetilde\theta_{t_k}$& TMLE & 87.1\%  & \textbf{477} & 386 & 595 & 1416& 1020 & 1938   \\ \\
      && \multicolumn{7}{ c }{$\theta=0$ (Null)}\\
  \cline{3-9} 
    Max. Sample Size & Unadj. & 5.10\% & \textbf{550} & 415 & 554  & 1808 & 1098 & 1831   \\
    with $\widehat\theta_{t_k}$& Stand. & 5.19\%  & \textbf{550} & 415 & 554 & 1807& 1098 & 1831   \\
    & TMLE & 5.18\%  & \textbf{549} & 415 & 554 & 1805& 1098 & 1830   \\
    Max. Sample Size & Stand. & 5.17\%  & \textbf{550} & 415 & 554 & 1807& 1098 & 1831   \\
    with $\widetilde\theta_{t_k}$& TMLE & 5.16\%  & \textbf{549} & 415 & 554  & 1805& 1098 & 1830  \\\\
    Information Adaptive & Unadj. & 5.33\%  & \textbf{628} & 459 & 634 & 2014& 1215 & 2042   \\
    with $\widehat\theta_{t_k}$& Stand. & 5.39\% & \textbf{569} & 434 & 574  & 1857& 1148 & 1882   \\
    & TMLE & 5.37\%  & \textbf{578} & 374 & 585 & 1881& 988 & 1910   \\
    Information Adaptive & Stand. & 5.38\% & \textbf{569} & 434 & 574   & 1857& 1148 & 1882  \\
    with $\widetilde\theta_{t_k}$& TMLE & 5.33\%  & \textbf{578} & 374 & 585 & 1881& 988 & 1910   \\\\ 
\hline
\end{tabular}
	{\raggedright Unadj., unadjusted estimator; Stand., standardization estimator; TMLE, targeted maximum likelihood estimator; ASN, average sample number; ASNj, average sample number at analysis $j$ ($j=1,2$); AAT, average analysis time (in days); AATj, average analysis time (in days) of analysis $j$ ($j=1,2$). \par}
\end{table}
\spacingset{1.85} 

\newpage

\section{Additional Simulation Study Under Violation of Independent Increments Property}\label{app:sec:sim}
In this section, we verify through simulations the operating characteristics of combining covariate adjustment (using the estimator described in Section \ref{sec:covAdj}) with group-sequential, information-adaptive designs under violation of the independent increments property. We present the simulation study as recommended in \cite{morris2019using}.

\subsection{Simulation Design}\label{app:subsec:sim_design}
\hspace{0.6cm}\textbf{Aims}:
To examine that the approach to combine covariate adjusted estimators with group sequential, information adaptive designs as explained in Section~\ref{sec:GSD} and Section~\ref{sec:inf} of the main article preserves (asymptotically) the Type I error. 
 
 \textbf{Data-Generating Mechanisms}: We consider two data-generating mechanisms for $(W, A, Y)$, both under a zero average treatment effect.
 \begin{enumerate}
     \item The vector of baseline covariates $W=(W_1, W_2, W_3, W_4)$ follows a standard multivariate normal distribution. The randomized treatment indicator $A$ follows a Bernoulli distribution with probability 0.5. The outcome $Y$ is measured 365 days after randomization. We generate $Y$ as $A\cdot Y(1)+(1-A)\cdot Y(0)$ where $Y(1)$ and $Y(0)$ follow a Bernoulli distribution with probability respectively equal to $expit(-2.5W_1+2W_2-2.5*W_3+2.1W_4)$ and $expit(-2W_1+2.5W_2-2.25*W_3-2.1W_4)$. This results in $P(Y=1|A=1)=0.5$ and $P(Y=1|A=0)=0.5$, and thus an average treatment effect $\theta=E\left(Y|A=1\right)-E\left(Y|A=0\right)$ of 0. 
\item The baseline covariate $W$ is normally distributed with mean 1 and standard deviation 1. The randomized treatment indicator $A$ follows a Bernoulli distribution with probability 0.5. The outcome $Y$ is measured 365 days after randomization. We generate $Y$ as $A\cdot Y(1)+(1-A)\cdot Y(0)$ where $Y(1)$ and $Y(0)$ follow a Bernoulli distribution with probability respectively $expit(W^2-\exp(W))$ and $expit(-\exp(W))$. This results in $P(Y=1|A=1)=0.17363$ and $P(Y=1|A=0)=0.11439$, and thus an average treatment effect $\theta=E\left(Y|A=1\right)-E\left(Y|A=0\right)$ of 0.05924 (i.e., the treatment effect under the alternative). Then, for each simulated participant with initial values $A=1$ and $Y=1$, we randomly replace $Y$ by an independent Bernoulli draw with probability of 0.6588 of being 1. This results in $P(Y=1|A=1)=0.17363\cdot 0.6588=0.11439$, and thus an average treatment effect $\theta$ of 0.
\end{enumerate} 

Interest lies in testing the null hypothesis $H_0: \theta= 0$ against the alternative $H_1: \theta \neq 0$, with $\theta$ defined as
$$\theta=E\left(Y|A=1\right)-E\left(Y|A=0\right),$$ 
at significance level $5\%$ with a power of 90\% under the alternative that $\theta_A=0.05$ for the first data-generating mechanism and $\theta_A=0.05925$ for the second data-generating mechanism. 

For a trial with two interim analyses at information fraction 0.50 and 0.75, the maximum information equals 
$$\mathcal{I}(\theta_A)=\left(\frac{z_{0.025}+z_{0.10}}{0.05}\right)^2\cdot IF=\left(\frac{1.96+1.28}{0.05}\right)^2\cdot 1.1553=4856, \text{ and}$$
$$\mathcal{I}(\theta_A)=\left(\frac{z_{0.025}+z_{0.10}}{0.05925}\right)^2\cdot IF=\left(\frac{1.96+1.28}{0.05924}\right)^2\cdot 1.1553=3459, $$
for respectively the first and second data-generating mechanism.
Here, the inflation factor was calculated for a group sequential design with 2 (efficacy) interim analysis at information fraction 0.50 and 0.75 with the \texttt{R} package \texttt{rpact}. 

A uniform recruitment rate of approximately 5 and 1 participants per day was considered for respectively the first and second data-generating mechanisms.
For the data generating mechanisms described above, we perform $100,000$ Monte Carlo runs.

\textbf{Targets:} Our target of interest is testing the null hypothesis of no (average) effect of treatment $A$ on the outcome $Y$ at a $5\%$ significance level.

\textbf{Methods of Analysis:}
For each of the data-generating mechanisms, the simulated trial dataset are analyzed using respectively the following methods:
\begin{enumerate}
	\item Test statistic based on the standardization estimator as described in Section \ref{sec:covAdj} of the Appendix with misspecified working models. To evaluate the operational characteristics under misspecified models, we will suppose that instead of being given the vector of baseline covariates $W$, the covariates actually seen by the data analyst are $W_1'=exp(W_1/2)$, $W_2'=W_2/(1+exp(W_1))+10$, $W_3'=(W_1W_3/25+0.6)^3$ and $W_4'=(W_2+W_4+20)^2$ \citep[see][]{kang2007demystifying}.
	Specifically, we fit a logistic regression model of $Y$ on $W_1'=exp(W_1/2)$ at the first interim analysis and on $W_2'=W_2/(1+exp(W_1))+10$ at the other analyses.
	\item Test statistic based on the standardization estimator as described in Section \ref{sec:covAdj} of the Appendix with misspecified working models. To evaluate the operational characteristics under misspecified models, we fit a logistic regression model of $Y$ on $W$ at the first interim analysis and on $|W|$ at the other analyses.
\end{enumerate}
Note that 2 different test statistics are evaluated as we also consider the `updated' version of the estimator following the approach described in Section~\ref{sec:GSD} of the main article.

Each simulated trial dataset is analyzed as a group sequential design with 2 interim analysis at information fraction 0.50 and 0.75. The boundaries are based on an error spending function that approximates Pocock boundaries. The `updated' version  of the estimators follow the approach described in Section~\ref{sec:GSD} of the main article. 
Finally, we assume that once recruitment has been stopped -because the projected $n_{max}$ is reached- it is not restarted again, even if the projection of $n_{max}$ seems to be too small.

\textbf{Performance Measures:} We assess the finite-sample (empirical) Type I error rate of the test of no treatment effect (w.r.t. the risk difference) on the considered outcome $Y$.

\subsection{Simulation Results}
Under the first data-generating mechanism, the empirical Type I error for the original sequence of estimators $\hat\theta_{t_k}$ was slightly inflated to $5.37\%$ (Monte Carlo error of approximately 0.0014). This is a consequence of a violation of the independent increments property. Specifically, we found that the correlation between $\hat\theta_{t_2}$ and $\hat\theta_{t_2}-\hat\theta_{t_1}$ is approximately equal to 0.16 and the correlation between $\hat\theta_{t_3}$ and $\hat\theta_{t_3}-\hat\theta_{t_1}$ is approximately equal to 0.13. Applying the proposal in Section~\ref{sec:GSD} of the main article seems to decrease the Type I error to 5.07\% (Monte Carlo error of approximately 0.0014), meaning that the Type I error is maintained. This is a consequence of the orthogonalization which ensures that the correlation between $\widetilde{\theta}_2$ and $\widetilde{\theta}_2-\widetilde{\theta}_1$ and the correlation between $\widetilde{\theta}_3$ and $\widetilde{\theta}_3-\widetilde{\theta}_1$ are not significantly different from zero (at the 1\% significance level).

The empirical Type I error for the original sequence of estimators $\hat\theta_{t_k}$ under the second data-generating mechanism was inflated to $5.51\%$ (Monte Carlo error of approximately 0.0014). This is a consequence of a violation of the independent increments property: we found that the correlation between $\hat\theta_{t_2}$ and $\hat\theta_{t_2}-\hat\theta_{t_1}$ is approximately equal to 0.11 and the correlation between $\hat\theta_{t_3}$ and $\hat\theta_{t_3}-\hat\theta_{t_1}$ is approximately equal to 0.09. The orthogonalization ensures that the correlation between $\widetilde{\theta}_2$ and $\widetilde{\theta}_2-\widetilde{\theta}_1$ and the correlation between $\widetilde{\theta}_3$ and $\widetilde{\theta}_3-\widetilde{\theta}_1$ are not significantly different from zero (at the 1\% significance level). Specifically, this leads to a decreased Type I error of 5.31\% (Monte Carlo error of approximately 0.0014). This small inflation in Type I error (after orthogonalization) is a small-sample problem and disappears for larger samples (i.e., higher information).

\bibliographystyle{Chicago}

\bibliography{sample.bib}
\end{document}